\documentclass[journal,draftclsnofoot,onecolumn,12pt]{IEEEtran}
\pdfoutput=1

\usepackage{amsthm,amssymb,graphicx,multirow,amsmath,color,amsfonts}
\usepackage[update,prepend]{epstopdf}
\usepackage[noadjust]{cite}
\usepackage[latin1]{inputenc}
\usepackage{tikz}
\usetikzlibrary{arrows,calc}		
\usepackage{bbm} 
\usepackage{pdfpages}
\usepackage{tabulary}
\usepackage{multirow}
\usepackage{comment}
\usepackage{mathtools}
\usepackage{bigints}

\def\nb0{{\mathbf{0}}}
\def\nb1{{\mathbf{1}}}

\def\nbbE{{\mathbb{E}}}

\def\nbbR{{\mathbb{R}}}

\newtheorem{lemma}{Lemma}
\newtheorem{thm}{Theorem}

\newtheorem{cor}{Corollary}

\newtheorem{remark}{Remark}

\def\E{\mathbb{E}}

\def\P{\mathbb{P}}
\def\pc{\mathtt{P_c}}

\def\T{\beta}							

\def\sir{\mathtt{SIR}}

\def\calE{\mathcal{E}}

\def\calL{\mathcal{L}}

\def\calC{\mathcal{C}}

\allowdisplaybreaks 

\begin{document}
\graphicspath{{./Figures/}}
\title{Coverage Analysis of a Vehicular Network Modeled as Cox Process Driven by Poisson Line Process }
\author{
Vishnu Vardhan Chetlur and Harpreet S. Dhillon
\thanks{The authors are with Wireless@VT, Department of ECE, Virginia Tech, Blacksburg, VA (email: \{vishnucr, hdhillon\}@vt.edu). The support of the US NSF (Grant IIS-1633363) is gratefully acknowledged.
\hfill Manuscript last updated: \today.}
}
\maketitle
\begin{abstract}
In this paper, we consider a vehicular network in which the wireless nodes are located on a system of roads. We model the roadways, which are predominantly straight and randomly oriented, by a Poisson line process (PLP) and the locations of nodes on each road as a homogeneous 1D Poisson point process (PPP). Assuming that each node transmits independently, the locations of transmitting and receiving nodes are given by two Cox processes driven by the same PLP. For this setup, we derive the coverage probability of a typical receiver, which is an arbitrarily chosen receiving node, assuming independent Nakagami-$m$ fading over all wireless channels. Assuming that the typical receiver connects to its closest transmitting node in the network, we first derive the distribution of the distance between the typical receiver and the serving node to characterize the desired signal power. We then characterize coverage probability for this setup, which involves two key technical challenges. First, we need to handle several cases as the serving node can possibly be located on any line in the network and the corresponding interference experienced at the typical receiver is different in each case. Second, conditioning on the serving node imposes constraints on the spatial configuration of lines, which require careful analysis of the conditional distribution of the lines. We address these challenges in order to accurately characterize the interference experienced at the typical receiver. We then derive an exact expression for coverage probability in terms of the derivative of Laplace transform of interference power distribution. We analyze the trends in coverage probability as a function of the network parameters: line density and node density. We also study the asymptotic behavior of this model and compare the coverage performance with that of a homogeneous 2D PPP model with the same node density.
\end{abstract}
\begin{IEEEkeywords}
	Stochastic geometry, Cox process, Poisson line process, coverage probability, vehicular network, road systems, Nakagami-$m$ fading.
\end{IEEEkeywords}

\section{Introduction} \label{sec:intro}
Vehicular communication, which collectively refers to vehicle-to-vehicle (V2V) and vehicle-to-infrastructure (V2I) communication, has enabled the vehicular nodes to share information with each other and also with roadside units (RSUs) to improve the road safety and transport efficiency~\cite{survey, traffic_safety, vc_its}. With autonomous vehicles becoming a reality in the near future, the data traffic originating from vehicular networks is expected to increase many folds while also putting more stringent latency and connectivity constraints compared to the networks of today. In order to meet these stringent requirements, it is critical to understand the system-level performance of these networks under different operational scenarios.
In the recent years, stochastic geometry has emerged as a powerful tool for modeling and system-level analysis of wireless networks. The most popular approach is to model the locations of wireless nodes by a homogeneous 2D Poisson point process (PPP) \cite{DhiGanBacAnd, jeff, tutorial_jeff, tutorial_hesham} and focus on the performance analysis of a randomly chosen receiver in the network. Despite its simplicity and analytical tractability, PPP may not always be a suitable model for all spatial configurations of nodes. In the context of this paper, the locations of vehicular nodes and RSUs in vehicular networks are restricted to roadways, which are predominantly linear and randomly oriented. The 2D PPP model, in which the location of nodes are modeled by randomly distributed points in the 2D plane, does not capture the coupling between the nodes and the underlying infrastructure (roads) in vehicular networks. While modeling the locations of vehicular nodes, one has to consider two fundamental sources of {\em randomness}: (i) the locations of nodes on each road are often irregular and can hence be treated as a realization of a point process, and (ii) the layout of the roads is also often irregular, which makes it possible to model the road system as a realization of a line process~\cite{robert,bacc_plp,volker1, volker, morlot}. In short, it is necessary to consider doubly stochastic spatial models for vehicular nodes that account for the randomness associated with the roads as well as the locations of nodes on these roads. A well-known {\em canonical} model in the literature that readily meets this requirement is a Cox process or doubly stochastic Poisson point process~\cite{stoyan, haenggi}, where the roads in a network are modeled by a Poisson line process (PLP) and the location of nodes on the roads are modeled by a 1D PPP. Despite the relevance of this canonical model in understanding the system-level performance of vehicular networks, its coverage analysis is still an open problem, which is the main focus of this paper. In particular, we develop tools to characterize serving distance as well as conditional interference power distributions, which collectively provide exact characterization of coverage probability and can also be readily applied to study many other aspects of vehicular networks. 

\subsection{Related Work} 
While there is a significant volume of literature pertaining to the analysis of vehicular networks using tools from stochastic geometry, the spatial models considered in these works are often too simple \cite{mit, busanelli, highway, mabiala, Ni, Steinmetz} and are limited to a single road or an intersection of two roads. For instance, a signal-to-interference plus noise ratio (SINR) based analysis to compute the optimum transmission probability for vehicles on a single road, has been proposed in \cite{mit}. The trade-offs between the aggregate packet progress and spatial frequency reuse for multi-hop transmission between vehicles in a multi-lane highway setup were studied in \cite{highway}. In~\cite{Ni,Steinmetz}, the authors have analyzed the packet reception probability of a link at the intersection of two perpendicular roads where the location of nodes are modeled as 1D PPP on each road. Since these models do not accurately capture the irregular structure of roads and their effect on the performance, they do not always offer reliable system-level insights that aid in the design. 

Although relatively sparse, there are also a few works in the literature where more sophisticated models that include the randomness associated with the road systems were studied~\cite{robert, bacc_plp, volker, volker1, volker2, multihop, morlot}. In \cite{robert}, the authors have modeled the streets in an urban setting by a Manhattan Poisson line process (MPLP) and the base stations on each road by a 1D PPP and characterized the downlink coverage performance of mmWave microcells by adopting a Manhattan distance based path-loss model. While this is a reasonable model for mmWave communication in an urban setting, it may not be applicable to all scenarios due to the irregular structure of roads. A more refined model for vehicular networks is presented in \cite{volker1, volker}, where the streets are modeled by the edges of either a Poisson-Line tessellation (PLT), Poisson-Voronoi tessellation (PVT), or a Poisson-Delaunay tessellation (PDT) and the nodes on each line are modeled by a homogeneous 1D PPP. Owing to its analytical tractability, PLT often gains preference over PVT and PDT in modeling road systems (it has also been used in other related applications, such as in modeling the effect of blockages in localization networks~\cite{aditya}). In \cite{volker1}, the authors have considered a hierarchical two-tier network whose components are modeled as a Cox process on a PLT and have characterized the mean shortest path on the streets connecting these components. Using the same spatial model, a formula for probability density function of inter-node distances was presented in \cite{volker}. In \cite{morlot}, the author has derived the uplink coverage probability for a setup where the typical receiver is randomly chosen from a PPP and the locations of transmitter nodes are modeled as a Cox process driven by a PLP. However, to the best of our knowledge, this paper is the first to derive the coverage probability for a setup where both the receiver and transmitter nodes are modeled by Cox processes driven by the same PLP. In other words, this paper is the first to derive the coverage probability of a vehicular node located on a PLP when it connects to another vehicular node on the same PLP. The technical challenges in this analysis originate from the spatial coupling between the vehicular nodes induced by the underlying PLP. More detailed account of our contributions is provided below. 

\subsection{Contributions}
In this paper, we present an analytical procedure for performing the canonical coverage analysis of a vehicular network. We consider a doubly-stochastic spatial model for wireless nodes, which captures the irregularity in the spatial layout of roads by modeling them as a PLP and the spatial irregularity in the locations of wireless nodes by modeling them as a 1D PPP on each road. In order to mimic various fading scenarios, we choose Nakagami-$m$ fading channel that allows us to control the severity of fading. For this setup, we derive the signal-to-interference ratio (SIR) based coverage probability of a typical receiver, which is an arbitrarily chosen receiving node in the network, assuming that it connects to its closest transmitting node in the network. We then study the trends in the coverage performance which offers useful system design insights. More technical details about the coverage probability and system-level insights are provided next. 

{\em Coverage probability.} We derive an exact expression for coverage probability by accurately characterizing the interference experienced at the typical receiver. We first derive several fundamental distance distributions that are necessary to characterize the desired signal power at the typical receiver. Since the distribution of nodes is coupled with the distribution of lines in the network, it poses two key challenges to the exact coverage analysis. First, the serving node, which is the closest transmitting node to the typical receiver, can possibly be located on any of the lines in the network. Consequently, the interference measured at the typical receiver in each of these cases is different and we have to handle each case separately. In order to address this issue, we derive a generalized expression for coverage probability for all these cases. Second, when a transmitting node on a particular line is chosen to be the serving node, it implies that there can not be any line with a node whose distance to the typical receiver is smaller than the distance between the typical receiver and the serving node. This additional constraint imposed by the distribution of nodes impacts the conditional distribution of lines as observed at the typical receiver. We determine the conditional distribution of the lines in order to accurately compute the interference at the typical receiver. We then determine the coverage probability in terms of derivative of Laplace transform of the distribution of the interference power.

{\em System-level insights.} Using our analytical results, we study the effect of two key network parameters, namely, node density and line density, on the coverage probability of the typical receiver. We observe that the coverage probability increases as the density of nodes on lines increases. However, the coverage probability degrades as the density of the lines in the network increases. The contrasting effect of node and line densities on the coverage probability offers useful insights in the design and deployment of RSUs in the network. We also compare the coverage probabilities of our setup with the results obtained from a homogeneous 2D PPP model, which is a widely accepted model for 2D wireless networks and is often used as a preferred approximation for more spohisticated point processes whose analysis may not be as tractable as a PPP. This comparison reveals that the 2D PPP model may not serve as a good approximation for our Cox process model, thereby highlighting the significance of our analytical results.

\section{Mathematical Preliminary: Poisson Line Process}
Since the PLP will be the main component of our model described in Section \ref{sec:sysmod}, a basic knowledge of its construction and properties will be useful in understanding the proposed model. While we provide only a brief introduction to PLP and its properties in this section, a detailed account of the underlying theory can be found in \cite{stoyan, schneider}.

\begin{figure}
	\centering
	\begin{minipage}[t]{.45\textwidth}
		\centering
		\includegraphics[scale=0.3]{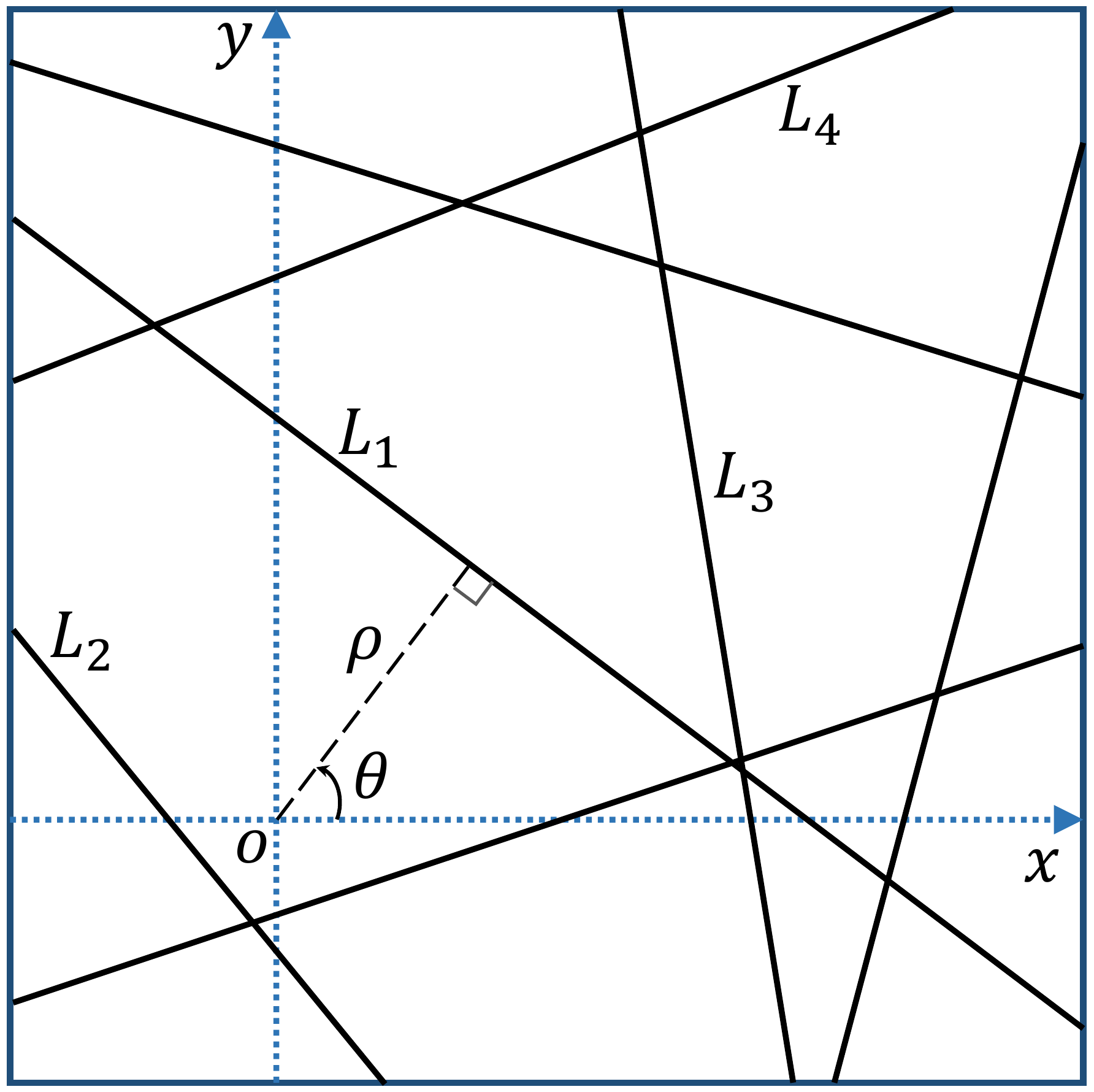}
		\caption{Illustration of Poisson line process in two-dimensional plane $\nbbR^2$.}
		\label{fig:prelim_r2}
	\end{minipage}%
	\quad 
	\begin{minipage}[t]{.45\textwidth}
		\centering
		\includegraphics[scale=0.3]{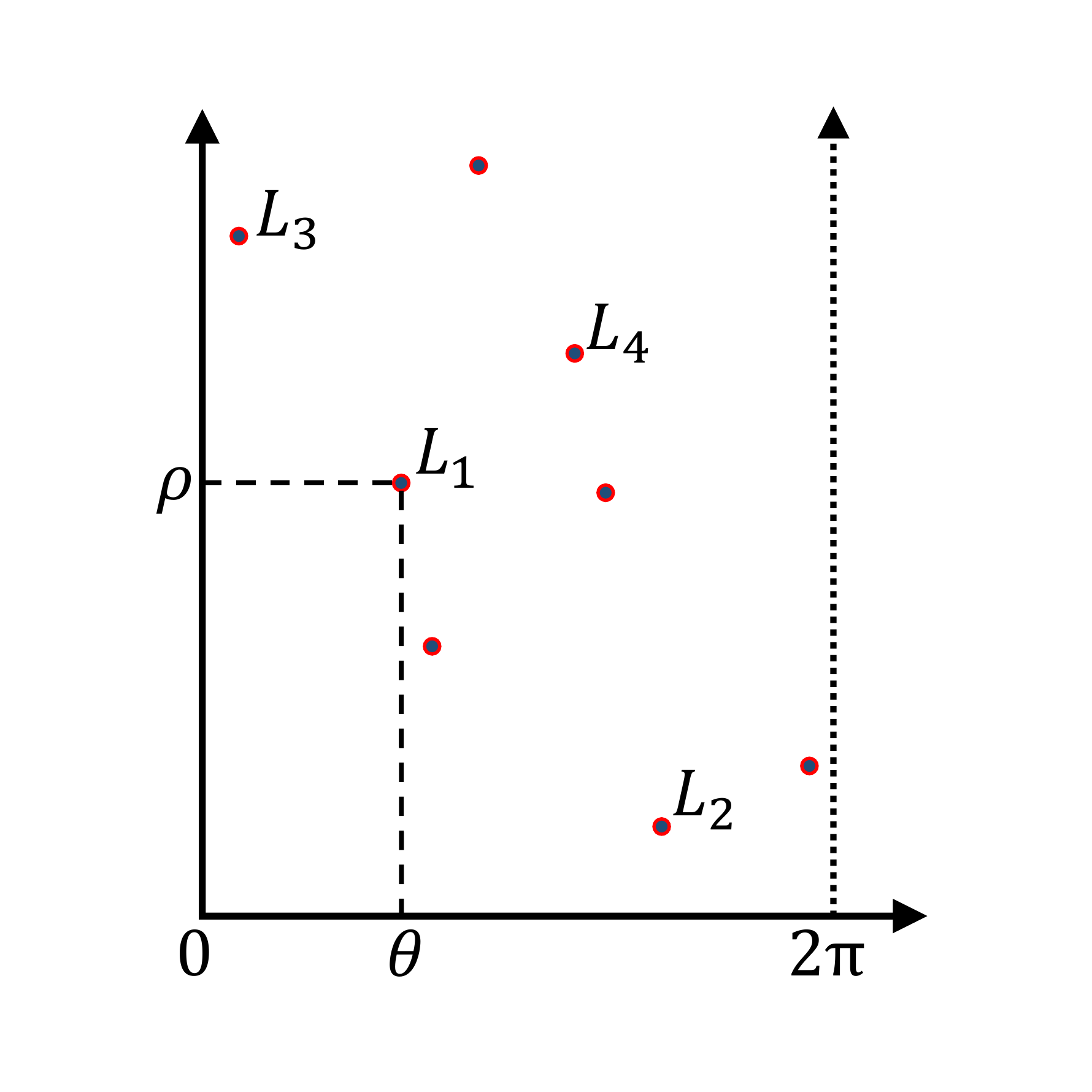}
		\caption{Illustration of a point process on representation space $\calC \equiv [0, 2 \pi) \times [0,\infty)$.}
		\label{fig:prelim_c}
	\end{minipage}
\end{figure}

{\em Line process}. A line process is simply a random collection of lines in a 2D plane. Any undirected line $L$ in $\nbbR^2$ can be uniquely characterized by its perpendicular distance $\rho$ from the origin $o \equiv (0,0)$ and the angle $\theta$ subtended by the perpendicular dropped onto the line from the origin with respect to the positive x-axis in counter clockwise direction, as shown in Fig. \ref{fig:prelim_r2}. The pair of parameters $\rho$ and $\theta$ can be represented as the coordinates of a point on the cylindrical surface $\calC \equiv [0, 2 \pi ) \times [0, \infty)$, which is termed as the {\em representation space}, as illustrated in Fig. \ref{fig:prelim_c}. Clearly, there is a one-to-one correspondence between the lines in $\nbbR^2$ and points on the cylindrical surface $\calC$. Thus, a random collection of lines can be constructed from a set of points on $\calC$. Such a set of lines generated by a Poisson point process on $\calC$ is called a Poisson line process. In our system model, we also assume the PLP to be motion-invariant for analytical simplicity. So, we will discuss the concept of motion-invariance for line processes and some well-established results of PLP next.

{\em Stationarity and Motion-Invariance.} The definition of stationarity for line processes is similar to that of point processes. A line process $\Phi_l = \{ L_1, L_2, \dots\}$ is said to be stationary if the translated line process $T\Phi_l = \{ T(L_1), T(L_2), \dots \}$ has the same distribution of lines as that of $\Phi_l$ for any translation $T$ in the plane. 
Upon translating the origin in the plane $\nbbR^2$ by a distance $t$ in a direction that makes an angle $\beta$ with respect to the positive x-axis, the equivalent representation of a line $L$ in $\calC$ changes from $(\rho, \theta)$ to $\big(\rho-t \cos(\theta - \beta), \theta\big)$. Therefore, for a stationary line process $\Phi_l$, the point process $\big\{\big(\rho_{L_1}-t \cos(\theta_{L_1} - \beta), \theta_{L_1}\big), \big(\rho_{L_2}-t \cos(\theta_{L_2} - \beta), \theta_{L_2}\big), \dots   \big\}$ in the representation space $\calC$ has the same distribution as that of the point process $\big\{\big(\rho_{L_1}, \theta_{L_1}\big), \big(\rho_{L_2}, \theta_{L_2}\big), \dots   \big\}$. Similarly, rotation of the axes about the origin by an angle $\gamma$ in $\nbbR^2$ changes the representation of the line in $\calC$ from $(\rho, \theta)$ to $(\rho, \theta - \gamma)$, where the operation $\theta-\gamma$ is modulo $2 \pi$. In addition to translation-invariance, if a line process is also invariant to the rotation of the axes about the origin, then it is said to be motion-invariant. 

{\em Line density.} Line density $\mu$ of a line process $\Phi_l$ is defined as the mean line length per unit area. If $\Phi_l$ is a motion-invariant line process, then the density of the corresponding point process $\lambda$ in the representation space $\calC$ is given by $\lambda = \frac{\mu}{ 2 \pi}$.

{\em Number of lines intersecting a disc.} If $\Phi_l$ is a motion-invariant Poisson line process with line density $\mu$, then the number of lines that intersect a convex region $K \subseteq \nbbR^2$ follows a Poisson distribution with mean
\begin{align}
\tau_K = \frac{\mu \nu(K) }{ 2 \pi} = \lambda \nu(K),
\end{align}  
where $\nu(K)$ is the perimeter of the convex region $K$. Therefore, the number of lines intersecting a disc of radius $d$ is Poisson distributed with mean $ 2 \pi \lambda d$.

\section{System Model} \label{sec:sysmod}

\subsection{Spatial Modeling of Wireless Nodes}
\begin{figure}
	\centering
	\includegraphics[scale=.3]{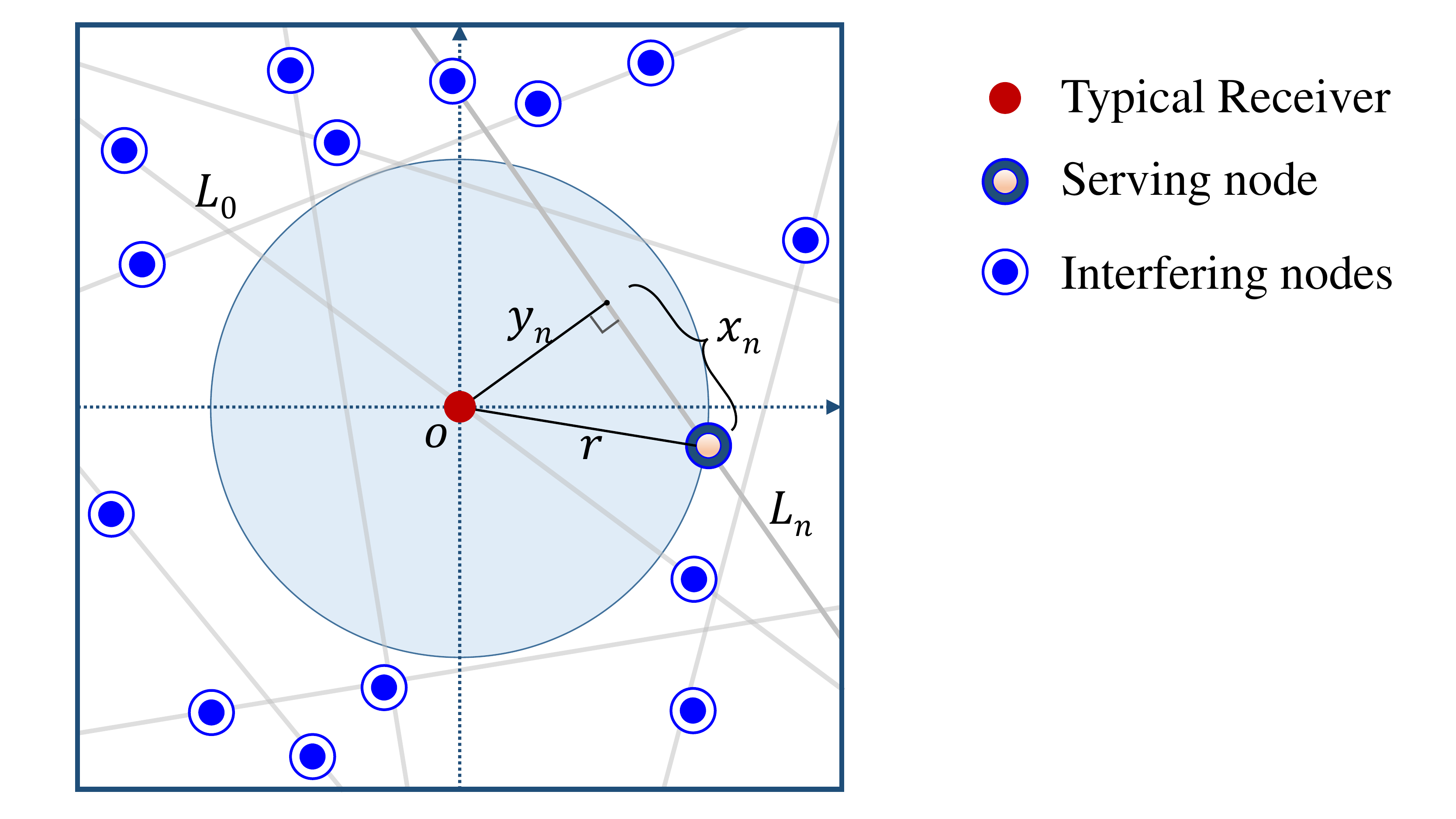}
	\caption{Illustration of the system model.}
	\label{fig:sysmod1}
\end{figure}

We first model the spatial distribution of road systems by a motion-invariant PLP $\Phi_l$ with line density $\mu_l$. We denote the density of equivalent PPP on the representation space $\calC$ by $\lambda_l$. We then model the locations of wireless nodes, which include vehicular nodes and RSUs, on each line (road) by a homogeneous 1D PPP with density $\lambda_n$. Assuming that each wireless node transmits independently with a probability $p$, the locations of transmitting nodes on each line is then given by a thinned PPP with density $\lambda_v = p \lambda_n$. We denote the set of locations of the transmitting nodes on a line $L$ by $\{ {\bf w}_L \} \equiv \Psi_L$. Similarly, the distribution of receiving nodes on each line is also a thinned PPP with density $\lambda_r = (1-p) \lambda_n$. Thus, the locations of transmitting and receiving nodes are modeled by Cox processes $\Phi_t$ and $\Phi_r$, which are driven by the same PLP $\Phi_l$. Our goal is to derive the SIR based coverage probability of a typical receiver from the point process $\Phi_r$. For analytical simplicity, we translate the origin $o \equiv (0,0)$ to the location of the typical receiver. The translated point process $\Phi_{r_0}$ can be treated as the superposition of the point process $\Phi_r$, an independent 1D PPP with density $\lambda_r$ on a line passing through the origin, and an atom at the origin $o$ \cite{morlot}. This can be understood by applying Slivnyak's theorem \cite{haenggi, stoyan} in two steps: first, we add a point at the origin to the PPP in the representation space $\calC$, thereby obtaining a PLP $\Phi_{l_0} =  \Phi_l \cup \{L_0\}$ with a line $L_0$ passing through the origin, and second, we add a point at the origin to the 1D-PPP on the line $L_0$ passing through the origin in $\nbbR^2$. The line $L_0$ passing through the origin will henceforth be referred to as the typical line. Since both $\Phi_r$ and $\Phi_t$ are driven by the same line process, the translated point process $\Phi_{t_0}$ is also the superposition of $\Phi_t$ and an independent PPP with density $\lambda_v$ on $L_0$, as shown in Fig. \ref{fig:sysmod1}. Since the other receiver nodes in the network do not have any impact on the SIR measured at the typical receiver in this setup, we will focus only on the distribution of transmitter nodes in the network. For brevity, the transmitter nodes will henceforth be referred to as only \textit{nodes}. We denote the $i^{th}\ (i = 1,2, \dots)$ closest line to the origin $o$ (excluding the typical line) by $L_i$ and its perpendicular distance to the origin by $Y_i$. The distance of the closest node on a line $L_i$ from the projection of the origin onto $L_i$ is denoted by $X_i$, as illustrated in Fig. \ref{fig:sysmod1}. Thus, the distance to the closest node on $L_i$ from the origin is $S_i = \sqrt{Y_i^2 + X_i^2}$. For notational consistency, we denote the distance of the typical line from the origin by $Y_0 \equiv 0$ and the distance to the closest node on $L_0$ by $S_0$. We denote the number of lines that intersect a region $A \subset \nbbR^2$ by $N_l (A)$ and the number of nodes in $A$ by $N_v(A)$. Throughout this paper, we denote the random variables by upper case letters and their corresponding realizations by lower case letters. For instance, $Y_n$ denotes a random variable, whereas $y_n$ denotes its realization.

\subsection{Transmitter Association Scheme and Propagation Model}
We assume that the transmit power is the same for all the nodes and the antennas are isotropic. We further assume that the typical receiver connects to its closest transmitting node in the network. Note that the closest node does not necessarily have to be on the same line as that of the typical receiver and can possibly be located on any of the other lines. We denote such an event in which the serving node is located on the $i^{th}$ closest line (excluding the typical line) to the origin by $\calE_i \ (i=1,2,\dots)$. We denote the event in which the serving node is located on the typical line by $\calE_0$.

In wireless communication networks, the severity of fading between the transmitter and the receiver depends on environmental factors and hence, the effect of fading can vary significantly from an urban scenario consisting of several buildings to rural areas and highways which are almost devoid of any tall structures. Therefore, in order to mimic a wide range of fading environments, we choose Nakagami-$m$ fading with parameter $m$. In the interest of analytical tractability, we restrict the values of $m$ to integers. For simplicity of exposition, we assume that the system is interference limited and hence, the thermal noise is neglected. Thus, the signal-to-interference ratio (SIR) at the typical receiver is
\begin{align}
\sir = \frac{G_0 R^{-\alpha}}{\sum_{L_j \in \Phi_{l_0}} \sum_{{\rm w}_{L_j} \in \Psi_{L_j} \setminus b(o, R)} G_{{\rm w}_{L_j}} \|{\rm w}_{L_j}\|^{-\alpha} },
\end{align}
where $\alpha > 2$ is the path-loss exponent, $G_0$ is the channel fading gain between the typical receiver and the serving node, $G_{{\rm w}_{L_j}}$ is the channel fading gain between the typical receiver and the interfering node at the location ${\rm w}_{L_j}$, $R$ is the Euclidean distance to the serving node from the typical receiver, and $\|{\rm w}_{L_j}\|$ is the Euclidean distance of the interfering node from the typical receiver.  

\section{Coverage Probability}
This is the main technical section of the paper, where we derive the coverage probability for the setup described in the previous section. Recall that the serving node which is the closest node to the typical receiver can possibly be located on any line $L_k \ (k=0,1, \dots)$. As a result, the interference at the typical receiver will be different in each of these cases and hence, they need to be handled separately. However, we can derive a generalized expression for the cases in which the serving node does not lie on the typical line ($\calE_1$, $\calE_2$, $\dots$). Therefore, in our analysis, we will derive the coverage probability conditioned on the events $\calE_0$ and $\calE_n \ (n = 1, 2, \dots)$ separately and obtain the final result using law of total probability. A key difference between the events $\calE_0$ and $\calE_n$ is that the distance of the line on which the serving node is located is always zero in case of $\calE_0$, whereas the distance of the line containing the serving node $Y_n$ in case of $\calE_n$ is a random variable. Therefore, in the computation of coverage probability conditioned on $\calE_n$, we will derive the intermediate results by additionally conditioning on $Y_n$. In the final step, we obtain the overall coverage probability by taking expectation over $Y_n$. While we can obtain some of the results for the case $\calE_0$ from the intermediate results pertaining to $\calE_n$ by simply substituting $Y_0 = 0$ in place of $Y_n$, we will provide detailed proofs for those results where this approach is not applicable.
 
\subsection{Preliminary Results}
We begin our analysis with the derivation of some fundamental distance distributions which will be used later in the computation of coverage probability. While it may be relatively straightforward to derive some of these results, they are presented here for completeness.

\begin{lemma}\label{lem:cdf_Yn}
	The cumulative distribution function (CDF) and probability density function (PDF) of the distance of the $n^{th}$ closest line from the origin $Y_n$ are 
	\begin{align}
	&\textrm{CDF:} \quad F_{Y_n} (y_n) = 1- \exp(-2 \pi \lambda_l y_n) \sum_{k=0}^{n-1} \frac{(2 \pi \lambda_l y_n)^k}{k!},\\\label{eq:fyn}
	&\textrm{PDF:} \quad f_{Y_n} (y_n) = \exp (-2 \pi \lambda_l y_n) \frac{(2 \pi \lambda_l y_n)^n}{y_n (n-1)!}.
	\end{align}
\end{lemma}
\begin{IEEEproof}
	From the definition of a PLP, recall that there is a one-to-one correspondence between lines in $\nbbR^2$ and points on $\calC \equiv [0 \  2\pi) \times [0 \ \infty) $. The abscissa and the ordinate of these points represent the orientation of the line and the distance of the line from the origin, respectively. We now consider the projections of these points onto the vertical axis of the cylindrical surface, which represents the distance of the lines from the origin. Note that the number of projections of points in a segment of length $t$ on the vertical axis of $\calC$ is the same as the number of points in the area $[0, 2\pi) \times [0 , t)$, which follows a Poisson distribution with mean $2 \pi \lambda_l t$. This means that the projections of points onto the vertical axis of $\calC$ forms a 1D PPP $\Psi_{l_0}$ with density $2 \pi \lambda_l$. Therefore, the distance of the $n^{th}$ closest line from the origin follows the same distribution as that of the distance of $n^{th}$ closest point in a 1D PPP with density $ 2 \pi  \lambda_l$, which is a well-known result in stochastic geometry \cite{haenggi}.
\end{IEEEproof}

\begin{lemma}\label{lem:cdfXn}
	Conditioned on the distance of the $n^{th}$ closest line to the origin $Y_n$, the CDF of the distance $X_n$ between the projection of origin onto the line $L_n$ and its closest node on $L_n$ is
	\begin{align}
	F_{X_n}(x_n|y_n) = 1- \exp (-2 \lambda_v x_n).
	\end{align}
\end{lemma}
\begin{IEEEproof}
	The proof follows from the void probability of a 1D PPP with density $\lambda_v$.
\end{IEEEproof}

\begin{lemma}\label{lem:cdfSn_Yn}
	Conditioned on the distance of the $n^{th}$ closest line to the origin $Y_n$, the CDF and PDF of the distance to the closest node on the line $L_n$ from the typical receiver  $S_n$ are 
	\begin{align}
	&\textrm{CDF:} \quad F_{S_n}(s_n|y_n) = 1- \exp (-2 \lambda_v \sqrt{s_n^2- y_n^2}),\\
	&\textrm{PDF:} \quad f_{S_n}(s_n| y_n) = \frac{2 \lambda_v s_n}{\sqrt{s_n^2-y_n^2}} \exp \big(-2 \lambda_v \sqrt{s_n^2-y_n^2}\big).
	\end{align}
\end{lemma}
\begin{IEEEproof}
	The conditional CDF of $S_n$ is given by 
	\begin{align*}
	F_{S_n}(s_n | y_n) &= \P ( S_n < s_n | Y_n )= \P ( \sqrt{X_n^2 + y_n^2} < s_n | Y_n) = \P (X_n < \sqrt{s_n^2-y_n^2} | Y_n) \\
	&= F_{X_n}(\sqrt{s_n^2-y_n^2}| y_n) = 1- \exp (-2 \lambda_v \sqrt{s_n^2- y_n^2}).
	\end{align*}
	The PDF $f_{S_n} (s_n | y_n)$ can be obtained by taking the derivative of $F_{S_n} (s_n|y_n)$ w.r.t. $s_n$.
\end{IEEEproof}

\begin{cor}\label{cor:cdfS0}
	The CDF and the PDF of the distance between the typical receiver at the origin and its closest node on the typical line $S_0$ are 
	\begin{align}
	\textrm{CDF:} \quad F_{S_0} (s_0) = 1 - \exp( - 2 \lambda_v s_0), \\
	\textrm{PDF:} \quad f_{S_0}(s_0) = 2 \lambda_v \exp(-2 \lambda_v s_0).
	\end{align}
\end{cor}
\begin{IEEEproof}
	The proof follows from substituting $S_0$ and $Y_0 = 0$ in place of $S_n$ and $Y_n$ in Lemma \ref{lem:cdfSn_Yn}.
\end{IEEEproof}
Conditioned on the distance $Y_n$, we will now derive the distribution of the distance of the closest node to the typical receiver among the nodes that are located on the lines that are closer and farther than the line of interest $L_n$ in the following Lemmas. These results will be used in the next subsection in the computation of the probability of occurrence of events $\calE_0$ and $\calE_n$.

\begin{lemma}\label{lem:cdf_Un_given_Yn}
	Conditioned on the distance of the $n^{th}$ closest line to the origin $Y_n$, the CDF and PDF of the distance $U_n$ between the typical receiver and its closest node among the $n-1$ lines $\{L_1, L_2, \ldots L_{n-1}\}$ (excluding the typical line) that are closer than $Y_n$ are  
	\begin{align}\notag
	&\textrm{CDF:}  \\
	&F_{U_n} (u_n | y_n) = \begin{dcases}
	1 -  \bigg(1 - \frac{u_n}{y_n} + \frac{1}{y_n} \int_{0}^{u_n} \exp\big(-\lambda_v 2 \sqrt{u_n^2 - z^2}\big) {\rm d}z \bigg)^{n-1}  , & 0  \leq u_n < y_n,\\ 
	1-\Bigg[ \bigg( \int_{0}^{y_n} \exp\big(-\lambda_v 2 \sqrt{u_n^2 - z^2}\big)\frac{{\rm d}z}{y_n}  \bigg)^{n-1} \Bigg],  & y_n \leq u_n < \infty,
	\end{dcases}\\ \notag
	&\textrm{PDF:}\\
	&f_{U_n} (u_n | y_n) = \begin{dcases}
	(n-1) \bigg(1 - \frac{u_n}{y_n} + \frac{1}{y_n} \int_{0}^{u_n} \exp\big(-\lambda_v 2 \sqrt{u_n^2 - z^2}\big) {\rm d}z \bigg)^{n-2} \\
	\qquad \qquad  \times \bigg( \int_0^{u_n} \exp (-2 \lambda_v \sqrt{u_n^2-z^2}) \frac{2 \lambda_v u_n}{y_n \sqrt{u_n^2-z^2}} {\rm d} z \bigg) , & 0  \leq u_n < y_n,\\	
	(n-1)\bigg( \int_{0}^{y_n} \exp\big(-\lambda_v 2 \sqrt{u_n^2 - z^2}\big)\frac{{\rm d}z}{y_n}  \bigg)^{n-2} \\
	\qquad \qquad \times \int_0^{y_n} \exp (-2 \lambda_v \sqrt{u_n^2-z^2}) \frac{2 \lambda_v u_n}{y_n \sqrt{u_n^2-z^2}} {\rm d} z,  & y_n \leq u_n < \infty.
	\end{dcases}
	\end{align}
\end{lemma}
\begin{IEEEproof}
	See Appendix \ref{app:cdfUn}.	
\end{IEEEproof}	

\begin{lemma}\label{lem:cdf_Vn_given_Yn}
	Conditioned on the distance of the $n^{th}$ closest line to the origin $Y_n$, the CDF and PDF of the distance $V_n$ between the typical receiver and its closest node among the lines \{$L_{n+1}$, $L_{n+2}$, $\dots$\} that are farther than $Y_n$  are
	\begin{align}\notag
	&\textrm{CDF:}\\
	&F_{V_n} (v_n | y_n) = 1 - \exp\Bigg[- 2 \pi \lambda_l \int_{y_n}^{v_n} \Big( 1 - \exp( -2 \lambda_v \sqrt{v_n^2 -z^2}) \Big) {\rm d} z  \Bigg], \quad y_n \leq v_n < \infty, \\\notag 
	&\textrm{PDF:}\\ \notag 
	&f_{V_n} (v_n | y_n) = 2 \pi \lambda_l \int_{y_n}^{v_n} \exp(-2 \lambda_v \sqrt{v_n^2-z^2}) \frac{2 \lambda_v v_n}{\sqrt{v_n^2 - z^2}} {\rm d}z \\ 
	& \qquad \qquad \qquad \times \exp\Bigg[- 2 \pi \lambda_l \int_{y_n}^{v_n} \Big( 1 - \exp( -2 \lambda_v \sqrt{v_n^2 -z^2}) \Big) {\rm d} z  \Bigg],  \quad y_n \leq v_n < \infty.
	\end{align}
\end{lemma}
\begin{IEEEproof}
See Appendix \ref{app:cdfVn}.	
\end{IEEEproof}

We can easily specialize the results of Lemma \ref{lem:cdf_Vn_given_Yn} to obtain the CDF and PDF of the distance between the typical receiver and its closest node among the lines that are farther than the typical line as given in the following Corollary. 

\begin{cor}\label{cor:cdfV0}
	The CDF and PDF of the distance $V_0$ between the typical receiver and its closest node among the lines that are farther than the typical line are
	\begin{align}
	&\textrm{CDF:} \quad F_{V_0} (v_0 ) = 1 - \exp\Bigg[- 2 \pi \lambda_l \int_{0}^{v_0} \Big( 1 - \exp( -2 \lambda_v \sqrt{v_0^2 -z^2}) \Big) {\rm d} z  \Bigg], \\\notag
	& \textrm{PDF:} \quad f_{V_0} (v_0) = 2 \pi \lambda_l \int_{0}^{v_0} \exp\Big(-2 \lambda_v \sqrt{v_0^2-z^2}\Big) \frac{2 \lambda_v v_0}{\sqrt{v_0^2 - z^2}} {\rm d}z \\ 
	& \qquad  \qquad \qquad \qquad \times \exp\Bigg[- 2 \pi \lambda_l \int_{0}^{v_0} \Big( 1 - \exp( -2 \lambda_v \sqrt{v_0^2 -z^2}) \Big) {\rm d} z  \Bigg].
	\end{align}
\end{cor}
\begin{IEEEproof}
	The proof follows from substituting $V_0$ and $Y_0 = 0$ in place of $V_n$ and $Y_n$ in Lemma \ref{lem:cdf_Vn_given_Yn}.
\end{IEEEproof}
\subsection{Probabilities of Events $\calE_n$ and $\calE_0$}
In this subsection, we will derive the probability with which the typical receiver connects to a node on the $n^{th}$ closest line to the origin conditioned on the distance of the line from the origin $Y_n$ and the probability with which the typical receiver connects to a node on the typical line. These intermediate results hold the key to the derivation of conditional serving distance distribution in the next subsection. 

\begin{lemma}\label{lem:pen}
	Conditioned on $Y_n$, the probability of occurrence of the event $\calE_n$ is
	\begin{align}\label{eq:pen}
	\P ( \calE_n | Y_n)  =  \int_0^{\infty} \Big(1-F_{S_0}(s_n) \Big) \Big(1- F_{U_n}(s_n|y_n)\Big) 	\Big(1- F_{V_n} (s_n|y_n)\Big)f_{S_n} (s_n | y_n) {\rm d}s_n,
	\end{align}
	where  ${F_{S_0}} (\cdot)$, $ {F_{U_n}} ( \cdot\ | y_n)$, $ {F_{V_n}} (\cdot| y_n)$, and $f_{S_n}(s_n | y_n)$ are given by Corollary \ref{cor:cdfS0}, Lemmas \ref{lem:cdf_Un_given_Yn}, \ref{lem:cdf_Vn_given_Yn}, and \ref{lem:cdfSn_Yn}, respectively.
\end{lemma}
\begin{IEEEproof}
The typical receiver connects to a node on the $n^{th}$ closest line if the distance to the closest node on the line $L_n$ is smaller than the distance to the closest node on any other line. In this case, we will group all the lines excluding the line of interest $L_n$ into 3 sets: (i) the typical line $L_0$, (ii) the lines that are closer than the line $L_n$ ($L_1$, $L_2$, $\dots$, $L_{n-1}$), and (iii) the lines that are farther than the line $L_n$ ($L_{n+1}$, $L_{n+2}$, $\dots$). The distance to the closest node on $L_n$ must be smaller than the distance to the closest node in each of these three sets, i.e., $S_n$ must be smaller than the minimum of $S_0, U_n$, and $V_n$. Thus, the conditional probability of occurrence of the event $\calE_n$ is computed as 
	\begin{align*}
	\P (\calE_n | Y_n) &= \P ( S_n < \min\{S_0, U_n, V_n\} | Y_n ) =\P ( S_n < S_0, S_n < U_n , S_n < V_n | Y_n ) \\
	&\stackrel{(a)}{=} \int_0^{\infty} \P (S_0> s_n|S_n, Y_n) \P (U_n> s_n | S_n, Y_n) \P (V_n> s_n |S_n, Y_n) f_{S_n} (s_n | y_n) {\rm d} s_n\\
	&\stackrel{(b)}{=} \int_0^{\infty} \Big(1-F_{S_0}(s_n) \Big) \Big(1- F_{U_n}(s_n|y_n)\Big) 	\Big(1- F_{V_n} (s_n|y_n)\Big)f_{S_n} (s_n | y_n) {\rm d}s_n,
	\end{align*}
	where (a) follows from the conditional independence of the variables $S_0$, $U_n$, and $V_n$, and (b) follows from the independence of the random variable $S_0$.
\end{IEEEproof}

\begin{cor}\label{cor:pe0}
	The probability of occurrence of the event $\calE_0$ is 
	\begin{align}\label{eq:pe0}
	\P (\calE_0) = \int_{0}^{\infty} F_{S_0}(v_0) f_{V_0}(v_0) {\rm d}v_0,
	\end{align}
	where $F_{S_0}(\cdot)$ and $f_{V_0}(\cdot)$ are given by Corollaries \ref{cor:cdfS0} and \ref{cor:cdfV0}, respectively.
\end{cor}
\begin{IEEEproof}
	The typical receiver at the origin connects to a node on the typical line when the distance to the closest node on the typical line is smaller than the distance to the closest node on any other line. Thus, the probability of occurrence of $\calE_0$ is given by
	\begin{align*}
	\P (\calE_0) = \P(S_0 < \min\{S_1, S_2, \ldots\}) = \P(S_0 < V_0) &= \int_{0}^{\infty} \P ( S_0 < v_0 | V_0) f_{V_0}(v_0) {\rm d} v_0\\
	&= \int_{0}^{\infty} F_{S_0}(v_0) f_{V_0}(v_0) {\rm d}v_0,
	\end{align*} 
	which completes the proof.
\end{IEEEproof}

\subsection{Serving Distance Distribution}
In this subsection, we will derive the distribution of distance between the typical receiver and the serving node conditioned on the events $\calE_n$ and $\calE_0$. As stated in the previous subsection, in case of $\calE_n$, the distance to the closest node on the $n^{th}$ closest line $S_n$ must be smaller than the minimum of $S_0, U_n$, and $V_n$. Therefore, we first determine the distribution of $W_n = \min\{S_0, U_n, V_n\}$ in the following Lemma.

\begin{lemma}\label{lem:cdfWn_Yn}
	Conditioned on the distance of the $n^{th} $ closest line to the origin $Y_n$, the CDF and PDF of $W_n =\min\{S_0, U_n, V_n\}$ are	
	\begin{align} 
	&\textrm{CDF:} \quad F_{W_n}(w_n| y_n) = 1 - \big(1 - {F_{S_0}} (w_n) \big) \big(1-{F_{U_n}} (w_n | y_n) \big) \big(1-{F_{V_n}} (w_n| y_n) \big),\\ \notag 
	&\textrm{PDF:} \quad f_{W_n} ( w_n | y_n) = {f_{S_0}} (w_n) \big(1-{F_{U_n}} (w_n | y_n) \big) \big(1-{F_{V_n}} (w_n| y_n) \big) + \big(1 - {F_{S_0}} (w_n) \big) {f_{U_n}} (w_n | y_n)  \\ 
	&\qquad \qquad \qquad \ \qquad\times \big(1-{F_{V_n}} (w_n| y_n) \big)  + \big(1 - {F_{S_0}} (w_n) \big) \big(1-{F_{U_n}} (w_n | y_n) \big) {f_{V_n}} (w_n| y_n) ,
	\end{align}
	where $	{F_{S_0}} (\cdot)$, $	{f_{S_0}} (\cdot)$ are given by Corollary \ref{cor:cdfS0}, $ {F_{U_n}} ( \cdot\ | y_n)$, $ {f_{U_n}} ( \cdot\ | y_n)$ are given by Lemma \ref{lem:cdf_Un_given_Yn}, and $ {F_{V_n}} (\cdot| y_n)$, $ {f_{V_n}} (\cdot| y_n)$ are given by Lemma \ref{lem:cdf_Vn_given_Yn}.
\end{lemma} 
\begin{IEEEproof}
	The proof simply follows from the distribution of minimum of three independent random variables \cite{garcia}.
\end{IEEEproof}

Using the intermediate results derived thus far, we will now derive the conditional distribution of the serving distance $R$ in the following Lemma.
\begin{lemma}\label{lem:cdfR_En}
	Conditioned on the event $\calE_n$ and the distance of the $n^{th}$ closest line $Y_n$, the CDF and PDF of the serving distance $R$ are 
	\begin{align}
	&\textrm{CDF:} \quad F_{R} (r|\calE_n, Y_n) = 1- \frac{1}{\P (\calE_n| Y_n)} \int_r^{\infty} \big(F_{S_n} (w_n|y_n) - F_{S_n}(r|y_n)\big) f_{W_n} (w_n|y_n) {\rm d} w_n ,\\
	&\textrm{PDF:} \quad f_R(r| \calE_n, y_n) = \frac{1}{\P (\calE_n | Y_n)} \int_r^{\infty} f_{S_n} (r| y_n) f_{W_n} (w_n | y_n) {\rm d }w_n,	
	\end{align}
	where $F_{S_n}(\cdot | y_n)$, $f_{S_n}(\cdot | y_n)$ are given by Lemma \ref{lem:cdfSn_Yn}, $\P (\calE_n | Y_n)$ and $f_{W_n}(\cdot | y_n)$ are given by Lemmas \ref{lem:pen} and \ref{lem:cdfWn_Yn}, respectively.
\end{lemma}
\begin{IEEEproof}
	The conditional CDF of the serving distance $R$ is computed as
	\begin{align*}
	F_{R}& (r|\calE_n, Y_n) \\
	&=  1 - \P (R>r | \calE_n, Y_n) = 1 - \frac{\P ( R > r, \calE_n | Y_n)}{\P (\calE_n | Y_n)} \stackrel{(a)}{=} 1 - \frac{\P (S_n > r, S_n < \min\{S_0, U_n, V_n\} | Y_n)}{\P (\calE_n | Y_n)} \\
	&= 1 - \frac{\P (r < S_n < W_n | Y_n)}{\P (\calE_n | Y_n)}  = 1- \frac{1}{\P (\calE_n|Y_n)} \int_r^{\infty} \P (r < S_n < w_n | W_n, Y_n) f_{W_n} (w_n|y_n) {\rm d} w_n\\
	&= 1- \frac{1}{\P (\calE_n| Y_n)} \int_r^{\infty} \big(F_{S_n} (w_n|y_n) - F_{S_n}(r|y_n)\big) f_{W_n} (w_n|y_n) {\rm d} w_n,
	\end{align*}
	where (a) follows from the condition for the occurrence of the event $\calE_n$. 
	The conditional PDF $f_R(r| \calE_n, y_n)$ can be computed by taking the derivative of $F_R(r| \calE_n, y_n)$ w.r.t. $r$.
\end{IEEEproof}

\begin{lemma}
	Conditioned on the event $\calE_0$, the CDF and PDF of the serving distance $R$ are given by
	\begin{align}
	&\textrm{CDF:} \quad F_{R} (r|\calE_0) = 1- \frac{1}{\P (\calE_0)} \int_r^{\infty} \big(F_{S_0} (v_0) - F_{S_0}(r)\big) f_{V_0} (v_0) {\rm d} v_0, \\
	&\textrm{PDF:} \quad f_R(r| \calE_0) = \frac{1}{\P (\calE_0)} \int_r^{\infty} f_{S_0}(r) f_{V_0} (v_0) {\rm d } v_0,
	\end{align}
where $F_{S_0}(\cdot)$, $f_{S_0}(\cdot)$ are given by Corollary \ref{cor:cdfS0}, $f_{V_0}(v_0)$ and $\P (\calE_0)$ are given by Corollaries \ref{cor:cdfV0} and \ref{cor:pe0}, respectively.
\end{lemma}
\begin{IEEEproof}
	The proof follows along the same lines as that of Lemma \ref{lem:cdfR_En}.
\end{IEEEproof}

\subsection{Conditional Probability Mass Function of Number of Lines}
\begin{figure}
	\centering
	\includegraphics[scale=.35]{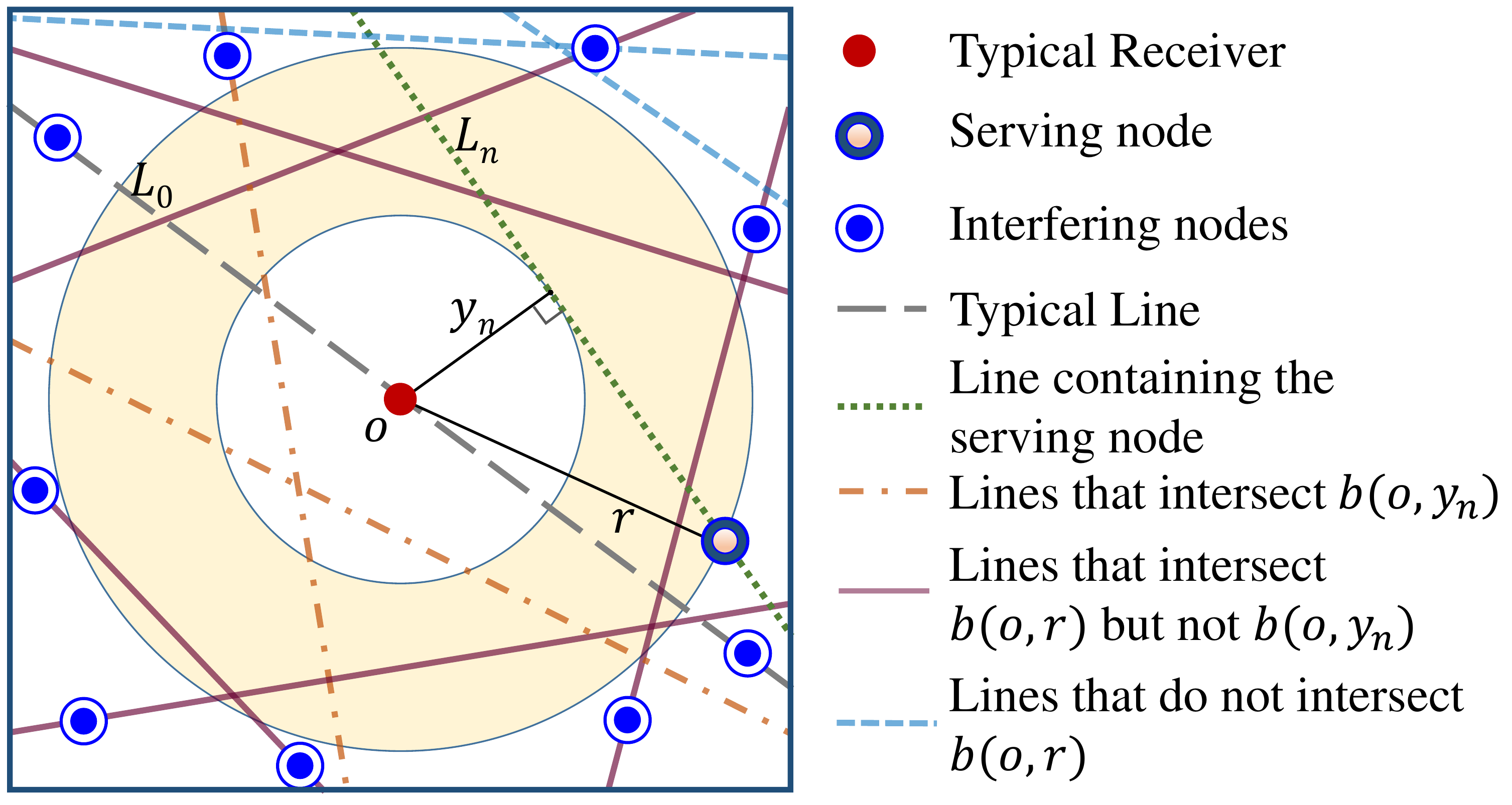}
	\caption{Illustration of different sets of lines.}
	\label{fig:pmf}
\end{figure}

Now that we have derived the distribution of the serving distance $R$, the next main step is to characterize the interference experienced at the typical receiver conditioned on $R$ for the events $\calE_n$ and $\calE_0$. The sources of interference are all the nodes that are located at a distance farther than $R$ from the origin, i.e., the nodes that lie outside the disc $b(o, R)$ centered at the origin $o$ with radius $R$. Please note that such nodes could also lie on lines which are located closer than $R$. Please see Fig. \ref{fig:pmf} for an illustration. Therefore, in order to accurately characterize the interference, we will have to first determine the distribution of lines. In order to explain this concretely, let us consider the case of $\calE_n$, where the serving node is located at a distance $R$ on the $n^{th}$ closest line from the origin. From the properties of PLP, we know that the number of lines intersecting a disc of fixed radius follows a Poisson distribution with mean equal to the line density scaled by the perimeter of the disc. However, this does not hold for the {\em conditional distribution} of the number of lines that intersect the disc $b(o,R)$. This is because the lines that intersect the disc $b(o, R)$ must not contain any nodes in the chord segment inside the disc (since we have already conditioned on the event that the serving node is located on $L_n$ at a distance $R$ from the typical receiver). This additional constraint imposed by the distribution of nodes on the lines impacts the conditional distribution of number of lines intersecting the disc $b(o,R)$. Note that the typical line is not included in the count of number of lines intersecting the disc $b(o, R)$. Now, conditioned on the event $\calE_n$, we know that there are at least $n$ lines that intersect the disc $b(o,R)$ which include the $n-1$ lines that are closer than the line of interest $L_n$ and the line $L_n$ itself which is at a distance $Y_n \leq R$. In addition to these $n$ lines, there are also a random number of lines that are farther than $Y_n$ but closer than the serving distance $R$, as illustrated in Fig. \ref{fig:pmf}. Therefore, our immediate goal is to determine the conditional distribution of this random number of lines that intersect the disc $b(o, R)$ but do not intersect the disc $b(o, Y_n)$, denoted by $N_l\big(b(o, R) \setminus b(o, Y_n) \big)$.

\begin{lemma}\label{lem:pmfNl_given_En}
	Conditioned on the event $\calE_n$, the distance of the $n^{th}$ closest line $Y_n$, and the serving distance $R$, the probability mass function (PMF) of number of lines that are farther than $Y_n$ and closer than $R$ is
	\begin{align}\notag
	\P \Big(N_l\big(b(o, r) \setminus b(o, y_n) \big)=k| \calE_n, Y_n, R\Big)&= \exp\Big(-2 \pi \lambda_l \int_{y_n}^{r} \exp(-2 \lambda_v \sqrt{r^2-y^2}) {\rm d}y \Big) \\ \label{eq:pmf_en}
	&\qquad \times \frac{\Big(2 \pi \lambda_l \int_{y_n}^{r} \exp(-2 \lambda_v \sqrt{r^2-y^2}) {\rm d}y \Big)^k}{k!} .	
	\end{align}
\end{lemma}
\begin{IEEEproof}
See Appendix \ref{app:pmf}.
\end{IEEEproof}

\begin{remark}
	Conditioned on the event $\calE_n$, serving distance $R$, and the distance of the $n^{th}$ closest line to the origin $Y_n$, we have derived the PMF of number of lines that are farther than $Y_n$ and closer than $R$. However, it can be observed that the conditional distribution remains Poisson but with a mean of $2 \pi \lambda_l  \int_{y_n}^{r} \exp(-2 \lambda_v \sqrt{r^2-y^2}) {\rm d}y $. Therefore, the conditional distribution of lines can be interpreted as a thinned Poisson line process with line density $  \frac{\mu_l}{2 \pi (r-y_n)}  \int_{y_n}^{r} \exp\big(-2 \lambda_v \sqrt{r^2-y^2}\big) {\rm d}y $. 
\end{remark}

\begin{cor}\label{cor:pmfNl_given_E0}
	Conditioned on the event $\calE_0$ and the serving distance $R$, the PMF of number of lines (excluding the typical line) that are closer than $R$ is
	\begin{align}\notag
	\P (N_l\big(b(o, R)\big)=k| R, \calE_0) &= \exp\Big(-2 \pi \lambda_l \int_0^{r} \exp(-2 \lambda_v \sqrt{r^2-y^2}) {\rm d}y \Big) \\
	&\qquad \times \frac{\Big(2 \pi \lambda_l \int_0^{r} \exp(-2 \lambda_v \sqrt{r^2-y^2}) {\rm d}y \Big)^k}{k!}.
	\end{align}
\end{cor}
\begin{IEEEproof}
	The proof follows from substituting $\calE_0$ and $Y_0 = 0$ in place of $\calE_n$ and $Y_n$ in Lemma \ref{lem:pmfNl_given_En}. 
\end{IEEEproof}

\subsection{Laplace Transform of Interference Distribution}
In this subsection, we will determine the Laplace transform of the distribution of the interference power conditioned on the serving distance $R$. We will first consider the case where the typical receiver connects to the $n^{th}$ closest line to the origin. In this case, we group the sources of interference into the following five sets: (i) the set of nodes present on the typical line, (ii) the set of nodes present on the line that contains the serving node which is at a distance $Y_n$, (iii) the set of nodes present on the lines that are closer than $Y_n$ (excluding the typical line), (iv) the set of nodes present on the lines that are farther than $Y_n$ but closer than the serving distance $R$ (an annular region), and (v) the set of nodes present on the lines whose distance from the origin exceeds the serving distance $R$. We denote the interference from these five sets of nodes by $I_0$, $I_{n}$, $I_{in}$, $I_{ann}$, and $I_{out}$, respectively. We will now derive the Laplace transform of distribution of interference from each of these components.

The interference measured at the typical receiver from the nodes on the typical line is given by $I_0 = \sum_{ {\rm w}_{L_0} \in \Psi_{L_0} \setminus b(o,R) } G_{{\rm w}_{L_0}} \| {\rm w}_{L_0} \|^{-\alpha}$, where $\Psi_{L_0}$ is the 1-D PPP on the typical line, $G_{{\rm w}_{L_0}}$ are the channel gains between the typical receiver and the interfering nodes at ${\rm w}_{L_0}$. While the Laplace transform of distribution of interference from nodes of a 1D PPP is very well-known, we still present the result in the following Lemma for completeness.
\begin{lemma}\label{lem:lapI0_En} 
	Conditioned on the event $\calE_n$, serving distance $R$, and the distance of the $n^{th} $ closest line $Y_n$, the Laplace transform of distribution of interference from the nodes situated on the typical line $L_0$ is 
	\begin{align}
	\calL_{I_0}(s | r, y_n, \calE_n) = \exp\Bigg[-2 \lambda_v \bigintssss_{r}^{\infty} \Big(1 -   \Big(1+ \frac{s x^{-\alpha}}{m} \Big)^{-m} \Big){\rm d}x \Bigg].
	\end{align}
\end{lemma}
\begin{IEEEproof}
	The distribution of nodes on the typical line is independent of the distance of the $n^{th}$ closest line $Y_n$. Thus, the Laplace transform of distribution of interference from nodes on the typical line can be computed as 
	\begin{align*}
	\calL_{I_0}&(s | r, y_n, \calE_n) = \E[e^{-sI_0}] =\E \  \E_G\Bigg[ \prod_{{\rm w}_{L_0} \in \Psi_{L_0} \setminus b(o, R)} \exp\big(-s G_{{\rm x}_{L_0}} \| {\rm w}_{L_0} \|^{-\alpha}\big) \Bigg] \\
	&\stackrel{(a)}{=}\E \Bigg[ \prod_{{\rm w}_{L_0} \in \Psi_{L_0} \setminus b(o,R)} \Big(1+ \frac{s \|{\rm w}_{L_0} \|^{-\alpha}}{m} \Big)^{-m} \Bigg] \stackrel{(b)}{=} \exp\Bigg[-2 \lambda_v \int_{r}^{\infty} \bigg(1 -  \Big(1+ \frac{s x^{-\alpha}}{m} \Big)^{-m} \bigg){\rm d} x \Bigg],
	\end{align*}
	where (a) follows from the Gamma distribution of channel fading gains, and (b) follows from the PGFL of PPP and substituting $x = \| {\rm w}_{L_0} \|$.  
\end{IEEEproof}

\begin{lemma}\label{lem:lapIn_En}
	Conditioned on the event $\calE_n$, the serving distance $R$, and the distance of the $n^{th}$ closest line form the origin $Y_n$, the Laplace transform of distribution of interference from the nodes on the line $L_n$ is
	\begin{align}
	\calL_{I_{n}}(s| r, y_n, \calE_n ) =  \exp\Bigg[-2 \lambda_v \bigintssss_{\sqrt{r^2 - y_n^2}}^{\infty} \bigg(1 -   \Big(1+ \frac{s (x^2 + y_n^2)^{-\alpha/2}}{m} \Big)^{-m} \bigg){\rm d}x \Bigg].
	\end{align}
\end{lemma}
\begin{IEEEproof}
	The proof follows along the same lines as that of Lemma \ref{lem:lapI0_En}.
\end{IEEEproof}

\begin{lemma}\label{lem:lap_Iin}
	Conditioned on the event $\calE_n$, serving distance $R$, and the distance of the $n^{th}$ closest line $Y_n$, the Laplace transform of distribution of interference from the nodes located on the lines that are closer than $Y_n$ is
	\begin{align}
	\calL_{I_{in} }(s|r, y_n, \calE_n) = \Bigg( \bigintsss_0^{y_n} \exp\Bigg[-2 \lambda_v \bigintssss_{\sqrt{r^2 - y^2}}^{\infty} \bigg(1 -   \Big(1+ \frac{s (x^2 + y^2)^{-\alpha/2}}{m} \Big)^{-m} \bigg){\rm d}x \Bigg] \frac{{\rm d}y}{y_n}\Bigg)^{n-1}.
	\end{align}
\end{lemma}
\begin{IEEEproof}
	See Appendix \ref{app:lap_Iin}.
\end{IEEEproof}

\begin{lemma}\label{lem:lapIin2_En}
	Conditioned on the event $\calE_n$, serving distance $R$, and the distance of the $n^{th}$ closest line from the origin $Y_n$, the Laplace transform of distribution of interference from the nodes located on the lines that are farther than $Y_n$ and closer than $R$ is
	\begin{align}\notag 
	\calL_{I_{ann}} (s| r, y_n, \calE_n) &= \exp\Bigg[\bigg(- 2 \pi \lambda_l \int_{y_n}^r\exp(-2 \lambda_v \sqrt{r^2-z^2}) {\rm d} z  \bigg)  \bigg(1- \bigintsss_{y_n}^{r} \exp\bigg[-2 \lambda_v \\ 
	& \quad \times \bigintssss_{\sqrt{r^2 - y^2}}^{\infty} \bigg(1 -   \Big(1+ \frac{s (x^2 + y^2)^{-\alpha/2}}{m} \Big)^{-m} \bigg){\rm d}x \bigg] \frac{{\rm d}y}{(r-y_n)} \bigg)\Bigg].
	\end{align}
\end{lemma}
\begin{IEEEproof}
	See Appendix \ref{app:lap_Iann}.
\end{IEEEproof}

\begin{lemma}\label{lem:lapIout_En}
	Conditioned on the event $\calE_n$, serving distance $R$, and the distance of the $n^{th}$ closest line from the origin $Y_n$, the Laplace transform of the distribution of the interference from the nodes located on the lines that are farther than the serving distance $R$ is 
	\begin{align}\notag
	\calL_{I_{out}} (s | r, y_n, \calE_n) &=  \exp\Bigg[ -2 \pi \lambda_l \bigintssss_r^{\infty} \Bigg( 1 - \exp\bigg[-2 \lambda_v \bigintssss_{0}^{\infty} \bigg(1 \\
	& \qquad \qquad \qquad  - \Big(1+ \frac{s (x^2 + y^2)^{-\alpha/2}}{m} \Big)^{-m} \bigg){\rm d}x \bigg] \Bigg) {\rm d}y \Bigg]
	\end{align}
\end{lemma} 
\begin{IEEEproof}
See Appendix \ref{app:lap_Iout}.
\end{IEEEproof}

The aggregate interference at the typical receiver is given by $ I = I_0 + I_{in} + I_n + I_{ann} + I_{out}$. Conditioned on $\calE_n$, $R$, and $Y_n$, the five components of interference are mutually independent and hence, the conditional Laplace transform of the distribution of total interference power is
\begin{align}\notag
\calL_I(s | r, y_n, \calE_n) &= \calL_{I_0}(s | r, y_n, \calE_n) \calL_{I_{in}}(s |r, y_n, \calE_n) \calL_{I_{n}}(s | r, y_n, \calE_n) \\ \label{eq:lapI}
& \qquad \qquad \times \calL_{I_{ann}}(s | r, y_n, \calE_n) \calL_{I_{out}}(s | r, y_n, \calE_n)  .
\end{align}  

In case of event $\calE_0$, the five different sources of interference mentioned earlier is reduced to three since the typical line and the line containing the serving node are the same and there are no lines closer than the typical line. Therefore, the sources of interference in this case are: (i) the set of nodes on the typical line, (ii) the set of nodes on the lines that are closer than the serving distance $R$, and (iii) the set of nodes on the lines that are farther than the serving distance $R$. We denote the interference from the three sets of nodes by $I_0$, $I_{in}$, and $I_{out}$, respectively. The conditional Laplace transform of distribution of the interference from these three sources can be directly obtained from the results in Lemmas \ref{lem:lapI0_En}, \ref{lem:lapIin2_En}, and \ref{lem:lapIout_En} by substituting $\calE_0$ and $Y_0 = 0$ in place of $\calE_n$ and $Y_n$, as given in the following Corollary.

\begin{cor}
	Conditioned on the event $\calE_0$ and the serving distance $R$, the Laplace transform of interference power distribution is 
	\begin{align}
	\calL_I (s| r, \calE_0) = \calL_{I_0} (s|r, \calE_0) \calL_{I_{in}} (s|r, \calE_0) \calL_{I_{out}} (s|r, \calE_0),
	\end{align}
	where
	\begin{align}
	\calL_{I_0}(s | r, \calE_0) &= \exp\Bigg[-2 \lambda_v \bigintssss_{r}^{\infty} \Big(1 -   \Big(1+ \frac{s x^{-\alpha}}{m} \Big)^{-m} \Big){\rm d}x \Bigg], \\ \notag 
	\calL_{I_{in}} (s| r, \calE_0) &= \exp\Bigg[\bigg(- 2 \pi \lambda_l \int_{0}^r\exp(-2 \lambda_v \sqrt{r^2-z^2}) {\rm d} z  \bigg)  \bigg(1- \bigintssss_{0}^{r} \exp\bigg[-2 \lambda_v \\ 
	& \qquad \times \bigintssss_{\sqrt{r^2 - y^2}}^{\infty} \bigg(1 -   \Big(1+ \frac{s (x^2 + y^2)^{-\alpha/2}}{m} \Big)^{-m} \bigg){\rm d}x \bigg] \frac{{\rm d}y}{r} \bigg)\Bigg],
	\end{align}
	and
	\begin{align} \notag
	\calL_{I_{out}} (s | r, \calE_0) &=  \exp\Bigg[ -2 \pi \lambda_l \bigintssss_r^{\infty} \Bigg( 1 - \exp\bigg[-2 \lambda_v \bigintssss_{0}^{\infty} \bigg(1 \\
	& \qquad \qquad \qquad  - \Big(1+ \frac{s (x^2 + y^2)^{-\alpha/2}}{m} \Big)^{-m} \bigg){\rm d}x \bigg] \Bigg) {\rm d}y \Bigg].
	\end{align}
\end{cor}

Now that we have determined the Laplace transform of distribution of interference power from all the components for both the cases $\calE_n$ and $\calE_0$, we will derive the coverage probability next.

\subsection{Coverage Probability}
The coverage probability is formally defined as the probability with which the SIR measured at the receiver exceeds a predetermined threshold $\beta$ required for a successful communication. Using the results derived thus far, the total coverage probability at the typical receiver can be obtained in terms of the conditional Laplace transform of the distribution of the interference power as given in the following theorem. 

\begin{thm}
	The coverage probability of the typical receiver $\pc$ is 
\begin{align}\notag
\pc &= \P (\calE_0) \sum_{k=0}^{m-1} \int_0^{\infty} \frac{(-m \T)^k}{ r^{- k \alpha} k!} \bigg[ \frac{\partial^k}{\partial s^k} \calL_I(s| r, \calE_0) \bigg]_{s=m \T r^{\alpha}} f_R(r|\calE_0) {\rm d}r  +  \sum_{n=1}^{\infty} \sum_{k=0}^{m-1}  \int_0^{\infty} \int_{y_n}^{\infty}  \frac{(-m \T )^k}{r^{-k \alpha}k!} \\ 
& \quad  \times \bigg[ \frac{\partial^k}{\partial s^k} \calL_I(s| r, \calE_n, y_n) \bigg]_{s=m \T r^{\alpha}} \P (\calE_n | Y_n) f_R(r|\calE_n, y_n)   f_{Y_n}(y_n) {\rm d}r {\rm d} y_n.
\end{align}
\end{thm}
\begin{IEEEproof}
	The coverage probability can be computed as
\begin{align} \notag 
\pc &= \P (\sir > \beta) = \sum_{i=0}^{\infty}  \P (\calE_i) \P (\sir > \beta | \calE_i) \\ \notag 
&= \P (\calE_0) \P (\sir > \beta | \calE_0)  + \sum_{n=1}^{\infty} \nbbE_{Y_n} \Big[ \P (\calE_n | Y_n) \P (\sir > \beta | \calE_n, Y_n) \Big] \\ \notag
&= \P (\calE_0) \nbbE_R \Big[ \P (\sir > \beta | \calE_0, R) \Big]  + \sum_{n=1}^{\infty} \nbbE_{Y_n} \bigg[ \P (\calE_n | Y_n) \nbbE_R \Big[ \P (\sir > \beta | \calE_n, R, Y_n)\Big]  \bigg]  \\ \notag
&= \P (\calE_0) \int_0^{\infty}  \P (\sir > \beta | \calE_0, R) f_R(r | \calE_0){\rm d}r  \\ \label{eq:pc_step1} 
&\qquad \qquad + \sum_{n=1}^{\infty} \int_0^{\infty} \int_{y_n}^{\infty} \P (\sir > \beta | \calE_n, R, Y_n) \P (\calE_n | Y_n) f_R(r|\calE_n, y_n) f_{Y_n}(y_n) {\rm d}r {\rm d}y_n
\end{align}
Following the same approach presented in~\cite{vishnuJ1, ralph, gen_fading}, we can obtain the final expression by rewriting the conditional coverage probability in \eqref{eq:pc_step1} in terms of derivative of Laplace transform of the distribution of the interference power. This completes the proof.
\end{IEEEproof}

\section{Numerical Results and Discussion}
In this section, we verify the accuracy of our analytical results by comparing the coverage probabilities evaluated using the theoretical expressions with the results obtained from the Monte-Carlo simulations. We also analyze the trends in coverage probability as a function of network parameters. We then highlight the significance of our analysis by comparing our coverage probability results with that of a homogeneous 2D PPP model with the same node density, which is often used to approximate more sophisticated point processes that may not be as tractable as a homogeneous PPP.

\subsection{Numerical Results}

\begin{figure}\centering
	\begin{minipage}[t]{.45\textwidth}
		\includegraphics[width=1\textwidth]{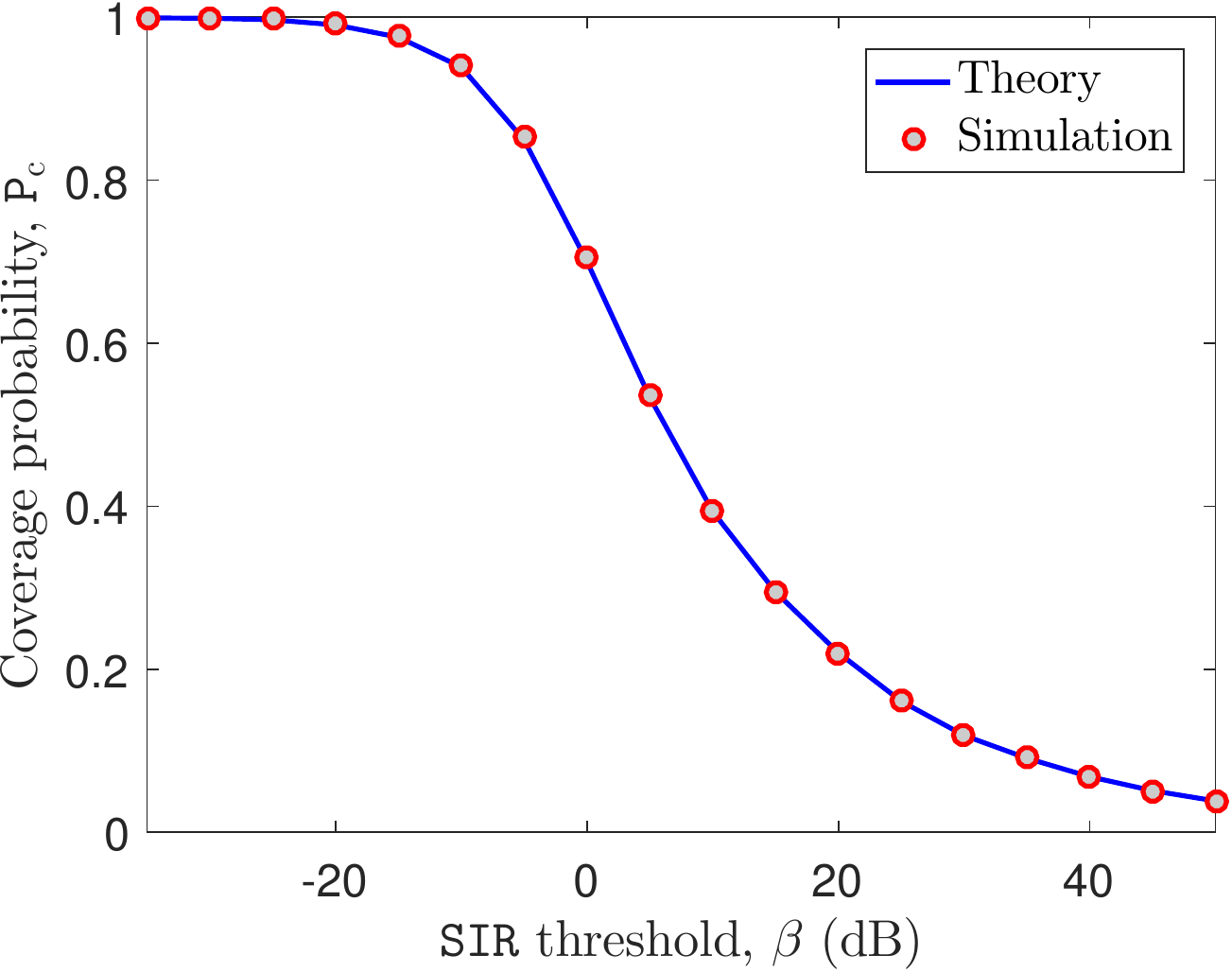}
		\caption{Coverage probability of the typical receiver as a function of $\sir$ threshold ($\mu_l=35$ km/km$^2$, $\lambda_v = 35$ nodes/km, $m = 1$, and $\alpha = 4$).}
		\label{fig:match}
	\end{minipage}%
	\qquad
	\begin{minipage}[t]{.45\textwidth}
		\includegraphics[width=1\textwidth]{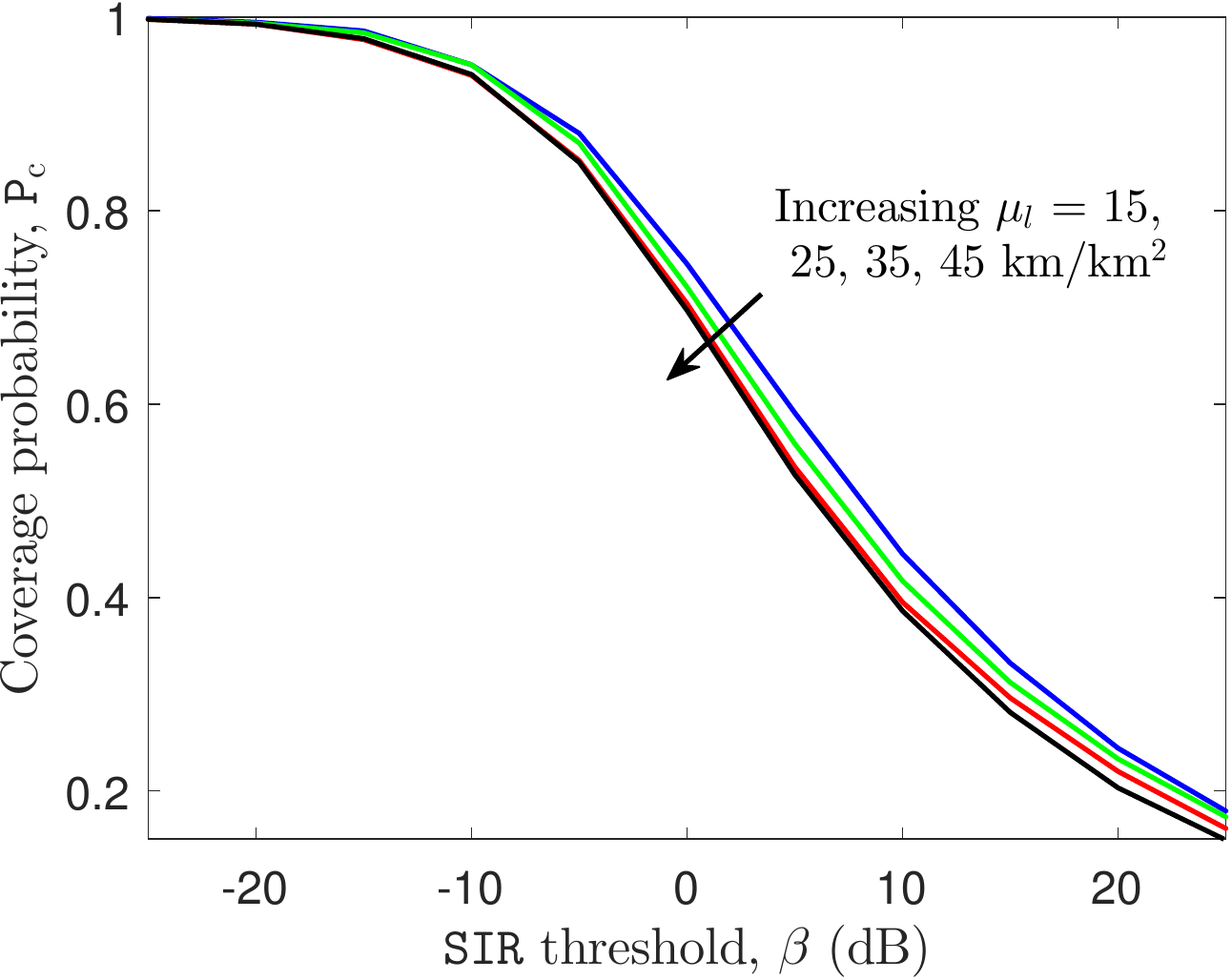}
		\caption{Coverage probability of the typical receiver as a function of $\sir$ threshold ($\lambda_v = 35$ nodes/km, $m = 1$, and $\alpha = 4$).}
		\label{fig:linedensity}
	\end{minipage}
\end{figure}

We simulate the Cox process model described in Section \ref{sec:sysmod} in MATLAB with line density $\mu_l = 35$ km/km$^2$, node density $\lambda_v = 35$ nodes/km, and path-loss exponent $\alpha = 4$. For this setup, we evaluate the empirical coverage probability for a receiver located at the origin using Monte-Carlo simulations. We observe that our analytical results match exactly with the Monte-Carlo simulations as depicted in Fig. \ref{fig:match}. The key network parameters that have an impact on the coverage probability are the line density and the node density. Therefore, we will next study the impact of each of these parameters separately on the coverage probability.

{\em Impact of line density.} We compute the coverage probability of the typical receiver for node density of $\lambda_v = 35$ nodes/km and different line densities of $\mu_l = 15, 25, 35,$ and $45$ km/km$^2$. It can be observed from Fig. \ref{fig:linedensity} that the coverage probability decreases as the line density increases. This trend in the coverage probability can be easily understood by examining the case where the receiver connects to a node on the typical line. In this case, an increase in the line density does not have any effect on the serving distance, however, it increases the interference power due to the reduced distance between the typical receiver and interfering nodes on other lines.

\begin{figure}\centering
	\begin{minipage}[t]{.45\textwidth}
		\includegraphics[width=1\textwidth]{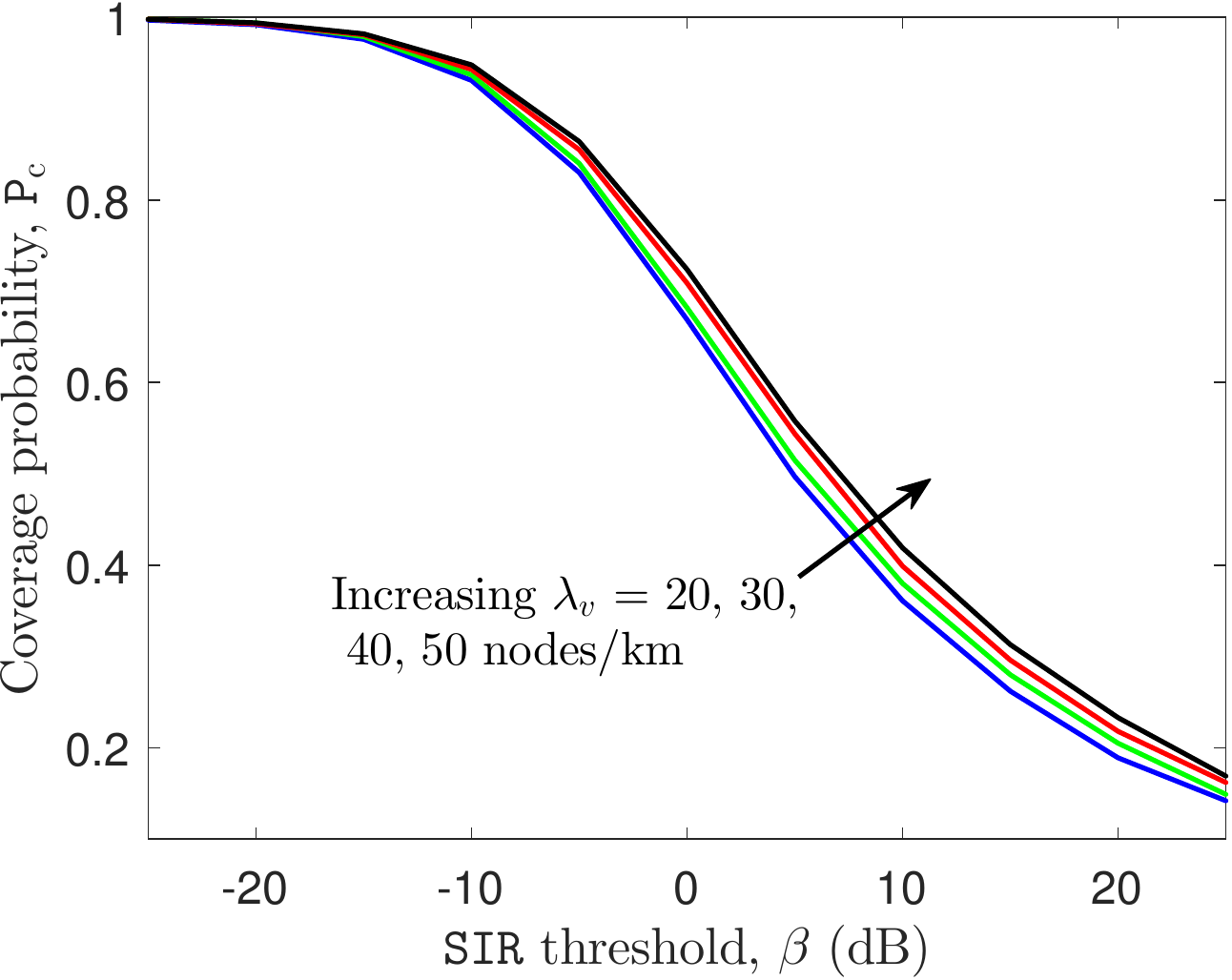}
		\caption{Coverage probability of the typical receiver as a function of $\sir$ threshold ($\mu_l = 35$ km/km$^2$, $m = 1$, and $\alpha = 4$).}
		\label{fig:nodedensity}
	\end{minipage}%
	\qquad
	\begin{minipage}[t]{.45\textwidth}
		\includegraphics[width=1\textwidth]{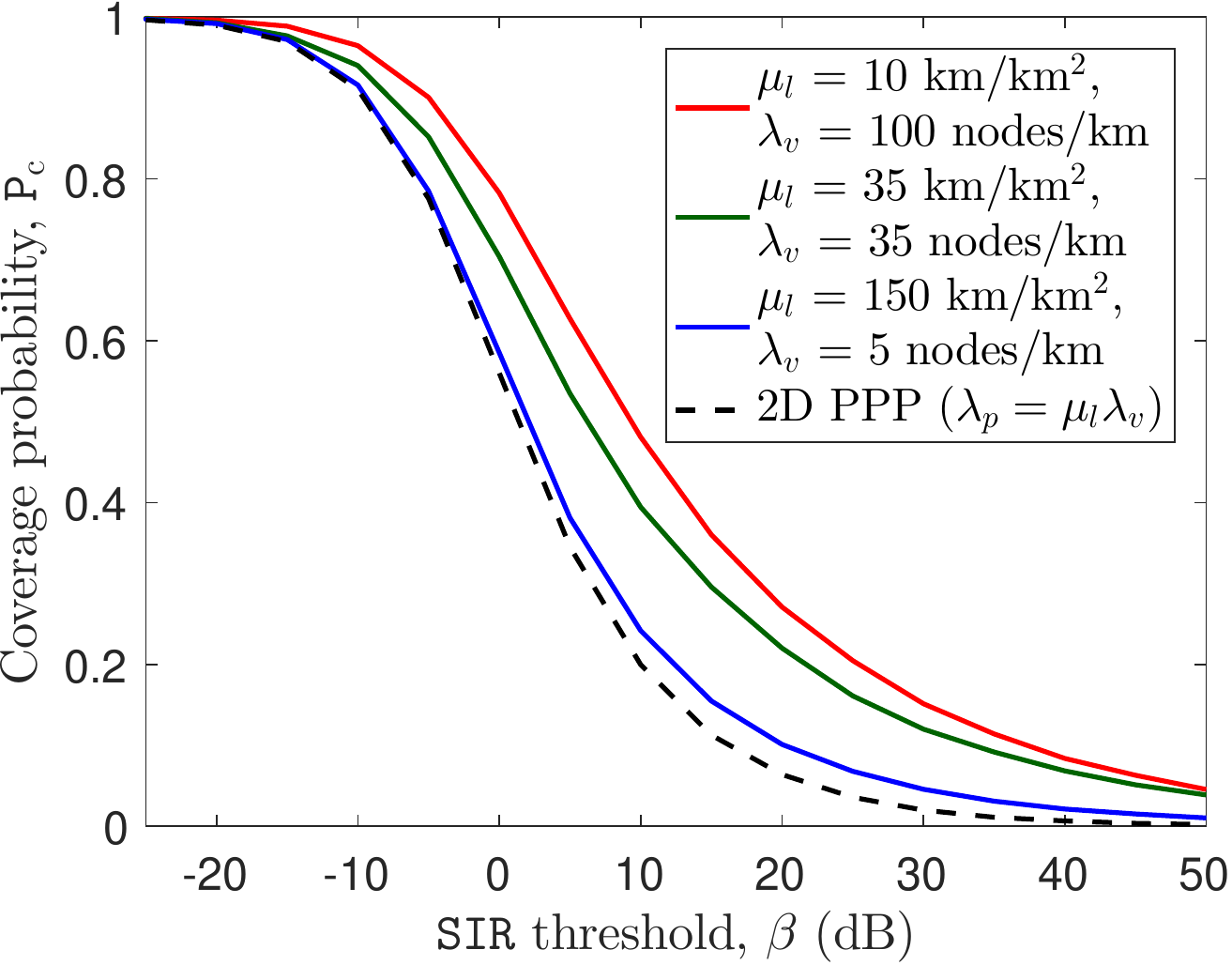}
		\caption{Coverage probability of the typical receiver as a function of $\sir$ threshold ($m=1$ and $\alpha = 4$).}
		\label{fig:ppp}
	\end{minipage}
\end{figure}

{\em Impact of node density.} We compare the coverage probability of the typical receiver for node densities of $\lambda_v = 20, 30, 40,$ and $50$ nodes/km as a function of $\sir$ threshold $\beta$. The other system parameters were fixed at $\mu_l = 35$ km/km$^2$, $m=1$, and $\alpha = 4$. It can be observed from Fig. \ref{fig:nodedensity} that the coverage probability increases as the density of nodes on the lines increases. Recall that the distance from the typical receiver at the origin to any node on a line involves two components: (i) perpendicular distance of the line from the origin, and (ii) the distance of the node (along the line) from the projection of the origin onto the line. When the density of nodes increases, the nodes come closer along the direction of the line, which decreases the second component of distance described above. Consequently, the decrement in the distance from the typical receiver to the nodes located on the lines that are closer to the origin is relatively more than the decrement in the distance to the nodes located on the lines that are farther away from the origin. This increases the desired signal power at a faster rate than the interference power, thus improving the SIR and hence the coverage probability at the typical receiver.  

\subsection{Comparison with a Homogeneous 2D PPP Model}
In this subsection, we compare the coverage probability of the typical receiver obtained for the Cox process driven by PLP described in Section \ref{sec:sysmod} with the results from a homogeneous 2D PPP model with the same node density, which is often used as a first-order approximation for more sophisticated point processes whose analysis may not be as tractable as a homogeneous PPP. We compute the coverage probability of the typical receiver for our setup with line density $\mu_l = 35$ km/km$^2$ and transmitter node density $\lambda_v = 35$ nodes/km. Thus, the equivalent node density of a homogeneous 2D PPP is $\lambda_p = \mu_l \lambda_v = 1225$ nodes/km$^2$. The notable difference between the coverage probabilities obtained from the two models in Fig. \ref{fig:ppp} highlights the importance of our model for the analysis of vehicular networks where the locations of nodes are restricted to road systems. This suggests that the homogeneous 2D PPP is not a good approximation for the proposed Cox process-based model in all operational scenarios. In order to glean better insights, we will now study the asymptotic behavior of the proposed model by evaluating the coverage probability for the extreme values of line and node densities.

{\em Low line density and high node density.} In this case, the collinearity of the locations of nodes in the network is more distinct. Hence, the coverage probability for our setup with $\mu_l = 10$ km/km$^2$ and $\lambda_v = 100$ nodes/km deviates significantly from the results of 2D PPP model, as shown in Fig. \ref{fig:ppp}. Also, due to the high density of nodes on each line, the typical receiver almost always connects to the closest node on the typical line. Therefore, the theoretical analysis for this case is relatively simple and the corresponding results are already given in the Corollaries.

{\em High line density and low node density.} As the density of lines $\mu_l \to \infty$, it allows the nodes to be positioned almost anywhere in $\nbbR^2$. In addition to this, if the density of nodes on each line is low, only a small number of nodes appear to be aligned on a straight line. As a result, the nodes appear to be uniformly distributed in the entire 2D plane like a PPP. This explains the trend in Fig. \ref{fig:ppp}, where the coverage probability of the typical receiver for our setup with a high line density of $\mu_l = 150$ km/km$^2$ and low node density of $\lambda_v = 5$ nodes/km is quite close to that of the 2D PPP model.

\section{Conclusion}
In this paper, we have presented an analytical method for the coverage analysis of a vehicular network in which the locations of nodes are confined to road systems. We have modeled the roads by a PLP and the nodes on each road by a homogeneous 1D PPP. Assuming that the typical receiver connects to its closest node in the network, we began our analysis with the derivation of several distance distributions which were necessary to determine the desired signal power at the typical receiver. We then computed the conditional distribution of lines in order to accurately characterize the interference at the typical receiver. We then derived an exact expression for SIR-based coverage probability of the typical receiver in terms of the derivative of Laplace transform of interference power distribution. We have verified the accuracy of our analytical results numerically by comparing them with the results obtained from Monte-Carlo simulations. Our analysis also reveals that the line density and the node density have a conflicting effect on the coverage probability. We then highlighted the significance of our analysis by comparing our coverage results with that of a homogeneous 2D PPP model with the same node density and also studied the asymptotic behavior of our model.

This work has numerous extensions. First and foremost, the proposed model as well as the canonical analysis presented in this paper can be readily applied to study other metrics of interest, such as information throughput and area spectral efficiency, which were not directly analyzed in this paper. The proposed approach can also be easily specialized to study the vehicular network performance under specific system constraints, such as those imposed by the mmWave frequencies, thus making the analysis relevant to a particular technology such as 5G. From stochastic geometry perspective, it will be useful to develop appropriate generative models for the proposed setup that simplify the analysis without compromising the accuracy of the results. Finally, while the proposed model is a reasonable canonical model for vehicular networks, there is always scope for making such models more accurate (often at the cost of reduced tractability) by obtaining network parameters from the actual data \cite{Gloaguen2006}. Therefore, another worthwhile extension of this work is to take a data-driven approach to vehicular network modeling, which will provide useful insights into the parameter ranges that are of interest in different morphologies. 

\appendix
\subsection{Proof of Lemma \ref{lem:cdf_Un_given_Yn} } \label{app:cdfUn}
	The CDF of $U_n$ conditioned on $Y_n$ is  
\begin{align*}
F_{U_n} (u_n | y_n) &= 1 - \P ( U_n > u_n | Y_n) = 1 - \P \Big( N_v\big(b(o,u_n)\big)=0 | Y_n\Big),
\end{align*}
where $N_v( b(o, u_n) )$ denotes the number of nodes within the disc of radius $u_n$ centered at $o$. Therefore, we need to find the probability that there are no nodes on any of the lines that are closer than $Y_n$ that intersect the disc $b(o, u_n)$. We know that there are $n-1$ lines whose distance from the origin is uniformly distributed in the range $(0, y_n)$. Depending on the range of $u_n$, there are two possible cases: (i) if $u_n$ is smaller than $y_n$, then the number of lines intersecting the disc $b(o,u_n)$ follows a binomial distribution with parameters $n-1$ and $\frac{u_n}{y_n}$, and (ii) if $u_n$ exceeds $y_n$, then all the $n-1$ lines intersect the disc $b(o,u_n)$. Thus, we obtain a piece-wise conditional CDF for $U_n$ as follows:	
\begin{align}\notag 
F_{U_n} &(u_n | y_n)  \\\notag 
&=\begin{dcases}
1 - \sum_{j=0}^{n-1} \P \Big( N_v\big(b(o,u_n)\big)=0 |  N_l\big(b(o,u_n)\big)=j , Y_n\Big) \\ 
\qquad  \times  \P \Big( N_l\big(b(o,u_n)\big)=j | Y_n\Big), & 0  \leq u_n < y_n,\\
1 - \prod_{i=1}^{n-1} \P \Big(N_v\big(L_i \cap b(o,u_n)\big) = 0 | Y_n \Big)  & y_n \leq u_n < \infty
\end{dcases}\\ \label{eq:cdfUn_step1}
&\stackrel{(a)}{=}  
\begin{dcases} 
1 - \sum_{j=0}^{n-1} \bigg( \P \Big( N_v \big(L \cap  b(o,u_n)\big) = 0 | Y_n\Big) \bigg)^j \P \Big( N_l\big(b(o,u_n)\big)=j | Y_n\Big), & 0  \leq u_n < y_n,\\
1 - \bigg[ \P \Big(N_v\big(L \cap b(o,u_n)\big) = 0 | Y_n \Big) \bigg]^{n-1} & y_n \leq u_n < \infty,
\end{dcases}
\end{align}
where (a) follows from the independent and identically distributed (i.i.d.) locations of the nodes on the lines. In step (a), $L$ denotes an arbitrarily chosen line that intersects the disc $b(o,u_n)$. We will now derive the expression for each term in \eqref{eq:cdfUn_step1}. For the case $0 \leq u_n < y_n$, we know that the number of lines intersecting the disc $b(o, u_n)$ follows a binomial distribution. Therefore, 
\begin{align}\label{eq:cdfUn_step1a}
\P \Big( N_l\big(b(o,u_n)\big)=j | Y_n\Big) = \binom{n-1}{j} \Big(\frac{u_n}{y_n}\Big)^j \Big(1 - \frac{u_n}{y_n}\Big)^{n-1-j}.
\end{align}
We will now evaluate the probability that there are no nodes on the segment of an arbitrarily chosen line $L$ that intersects the disc $b(o,u_n)$. We denote the perpendicular distance of the line from the origin by $Z$ which is uniformly distributed in the range $(0, u_n)$ for the first case. Conditioned on $Z$, the probability that there are no nodes on the segment that intersects $b(o, u_n)$ is given by 
\begin{align*}
\P \Big( N_v \big(L \cap  b(o,u_n)\big) = 0 | Z, Y_n\Big) = \exp\big(-\lambda_v 2 \sqrt{u_n^2 - z^2}\big). 
\end{align*}
This result follows from the void probability of 1D-PPP. By taking the expectation over $Z$, we now obtain the probability that there are no nodes on the segment of a randomly chosen line that intersects the disc $b(o,u_n)$ as
\begin{align}\label{eq:cdfUn_step1b}
\P \Big( N_v \big(L \cap  b(o,u_n)\big) = 0 | Y_n\Big) = \int_0^{u_n} \exp\big(-\lambda_v 2 \sqrt{u_n^2 - z^2}\big) \frac{1}{u_n} {\rm d} z. 
\end{align}

Similarly, for the second case where $y_n \leq u_n < \infty$, the distances of the lines that intersect the disc $b(0,u_n)$ are uniformly distributed in the range $(0, y_n)$. Therefore, the desired probability in this case is given by
\begin{align}\label{eq:cdfUn_step1c}
\P \Big( N_v \big(L \cap  b(o,u_n)\big) = 0 | Y_n\Big) = \int_0^{y_n} \exp\big(-\lambda_v 2 \sqrt{u_n^2 - z^2}\big) \frac{1}{y_n} {\rm d} z. 
\end{align}

Upon substituting \eqref{eq:cdfUn_step1a}, \eqref{eq:cdfUn_step1b}, and \eqref{eq:cdfUn_step1c} in \eqref{eq:cdfUn_step1} and simplifying the resulting expression, we obtain the final result. The PDF of $U_n$ can then be computed by taking the derivative of $F_{U_n}(u_n|y_n)$ with respect to $u_n$.

\subsection{Proof of Lemma \ref{lem:cdf_Vn_given_Yn} }\label{app:cdfVn}
The CDF of $V_n$ conditioned on $Y_n$ is  
\begin{align}\notag
&F_{V_n} (v_n | y_n) = 1 - \P ( V_n > v_n | Y_n) = 1 - \P \Big( N_v\big(b(o,v_n) \setminus b(o,y_n) \big)=0 | Y_n \Big) \\ \label{eq:cdfVn_step1}
&\stackrel{(a)}{=} 1 -  \sum_{n_l = 0}^{\infty} \P \Big(N_l\big(b(o, v_n) \setminus b(o,y_n)\big) = n_l | Y_n\Big) \bigg(  \P \Big(N_v\big(L \cap \{ b(o,v_n) \setminus b(o,y_n) \}\big)= 0 | Y_n\Big)  \bigg)^{n_l},
\end{align}
where (a) follows from the i.i.d. locations of nodes on the lines. Note that $L$ denotes an arbitrarily chosen line whose distance from the origin is greater than $v_n$ and smaller than $y_n$. From the definition of PLP, we know that the number of lines whose distances from the origin are in the range $(y_n, v_n)$ follows a Poisson distribution with mean $2 \pi \lambda_l (v_n - y_n)$. Thus,
\begin{align}\label{eq:cdfVn_step1a}
\P \Big(N_l\big(b(o, v_n) \setminus b(o,y_n)\big) = n_l | Y_n\Big) = \frac{\exp\big(-2 \pi \lambda_l (v_n-y_n)\big)\big(2 \pi \lambda_l (v_n-y_n) \big)^{n_l}}{n_l !}.
\end{align}
The evaluation of the second term in the equation \eqref{eq:cdfVn_step1} is similar to that of \eqref{eq:cdfUn_step1b} in the proof of Lemma \ref{lem:cdf_Un_given_Yn}. The only change is that the distances of the lines in this case are uniformly distributed in the range $(y_n, v_n)$. Hence, the desired probability is obtained as 
\begin{align}\label{eq:cdfVn_step1b}
\P \Big(N_v\big(L \cap \{ b(o,v_n) \setminus b(o,y_n) \}\big)= 0 | Y_n\Big) =  \int_{y_n}^{v_n} \exp\big(-\lambda_v 2 \sqrt{v_n^2 -z^2}\big)\frac{1}{(v_n-y_n)} {\rm d}z.
\end{align}
Substituting \eqref{eq:cdfVn_step1a} and \eqref{eq:cdfVn_step1b} in \eqref{eq:cdfVn_step1}, we obtain the final expression. The conditional PDF of $V_n$ can then be obtained by taking the derivative of $F_{V_n}(v_n|y_n)$  w.r.t. $v_n$.

\subsection{Proof of Lemma \ref{lem:pmfNl_given_En}} \label{app:pmf}
	The conditional PMF of number of lines can be computed as 
\begin{align}\notag
\P \Big( N_l\big(b(o, r) \setminus b(o, y_n) \big) = k \big| \calE_n, Y_n, R\Big) 	&\stackrel{(a)}{=} \frac{\P \Big( \calE_n \big| N_l \big( b(o, r) \setminus b(o, y_n) \big) = k, Y_n, R \Big)  }{\P \Big( \calE_n | Y_n, R\Big)} \\ \label{eq:pnl_ren}
&\qquad \qquad \times \P \Big( N_l \big( b(o, r) \setminus b(o, y_n) \big) = k | Y_n, R\Big),
\end{align} 
where (a) follows from the application of Bayes' theorem.
We now need to determine each term in \eqref{eq:pnl_ren} to compute the desired conditional PMF. The first term in the numerator is nothing but the probability that there are no nodes inside a disc of radius $r$ centered at the origin, given that there are $k$ lines that are farther than $y_n$ and closer than $r$. In addition to these $k$ lines, there are $n-1$ lines that are closer than $y_n$, one line at a distance of $y_n$, and the typical line that intersect the disc $b(o,r)$. Therefore, the probability that there are no nodes on any of these lines that intersect the disc $b(o,r)$ can be computed as follows:
\begin{align}\notag 
\P \Big( &\calE_n \big| N_l \big( b(o, r) \setminus b(o, y_n) \big) = k, Y_n, R \Big) \\ \notag 
&= \P \Big( N_v\big(b(o,r)\big) = 0 \big| N_l \big( b(o, r) \setminus b(o, y_n) \big) = k, Y_n, R  \Big) \\ \notag
& =  \P \Big(N_v\big(L_0 \cap b(o,r)\big) = 0 | Y_n, R\Big) \bigg[  \prod_{i=1}^{n-1} \P \Big(N_v \big( L_i \cap b(o,r) \big) = 0 | Y_n, R\Big) \bigg] \\ 
&\qquad \times  \P \Big(N_v \big(L_n \cap  b(o,r)\big) = 0 | Y_n, R \Big)  \bigg[\prod_{j=1 }^k \P \Big(N_v \big( L_{n+j} \cap b(o,r) \big) = 0 | Y_n, R\Big)\bigg]\\ \notag
&= \exp(-2 \lambda_v r) \bigg[ \int_0^{y_n} \exp(-2 \lambda_v \sqrt{r^2 - y^2}) \frac{{\rm d}y }{y_n} \bigg]^{n-1} \exp\big(-2 \lambda_v \sqrt{r^2 - y_n^2}\big) \\ \label{eq:pmf_num1}
& \qquad \times \bigg[ \int_{y_n}^r \exp\big(-2 \lambda_v \sqrt{r^2 - y^2}\big) \frac{{\rm d}y }{(r-y_n)} \bigg]^{k}.
\end{align}
The second term in the numerator in \eqref{eq:pnl_ren} is the probability that there are $k$ lines that are farther than $y_n$ and closer than $r$. Since both $r$ and $y_n$ are fixed, the number of lines that are farther than $y_n$ and closer than $r$ simply follows a Poisson distribution with mean $2 \pi \lambda_l (r-y_n)$. Therefore, the second term in the  numerator is given by 
\begin{align} \label{eq:pmf_num2}
\P \Big(N_l \big( b(o, r) \setminus b(o, y_n) \big) = k | Y_n, R\Big)  = \frac{ \exp\big( -2 \pi \lambda_l (r-y_n)\big) \big(2 \pi \lambda_l (r-y_n) \big)^k}{k!}.
\end{align} 
The denominator in \eqref{eq:pnl_ren} is the probability of occurrence of event $\calE_n$ conditioned on both $Y_n$ and $R$. This is nothing but the probability that there are no nodes inside the disc of radius $r$. This can be easily computed using law of total probability as follows:
\begin{flalign}\notag 
&\P \Big( \calE_n | Y_n, R\Big) = \sum_{k=0}^{\infty} \P \Big( \calE_n \big| N_l \big( b(o, r) \setminus b(o, y_n) \big) = k,  Y_n, R\Big) \P \Big(N_l \big( b(o, r) \setminus b(o, y_n) \big) = k | Y_n, R\Big) \\ \notag 
&\stackrel{(a)}{=} \exp(-2 \lambda_v r) \bigg[ \int_0^{y_n} \exp(-2 \lambda_v \sqrt{r^2 - y^2}) \frac{{\rm d}y }{y_n} \bigg]^{n-1} \exp\big(-2 \lambda_v \sqrt{r^2 - y_n^2}\big)   \\ \label{eq:pmf_den1}
& \quad \times  \exp\big( -2 \pi \lambda_l (r-y_n)\big) \sum_{k=0}^{\infty} \frac{\big(2 \pi \lambda_l (r-y_n) \big)^k}{k!} \bigg[ \int_{y_n}^r \exp\big(-2 \lambda_v \sqrt{r^2 - y^2}\big) \frac{{\rm d}y }{(r-y_n)} \bigg]^{k} \\ \notag
&= \exp \bigg(-2 \lambda_v r - 2 \lambda_v \sqrt{r^2-y_n^2} -2 \pi \lambda_l \int_{y_n}^r 1 - \exp \big(- 2 \lambda_v \sqrt{r^2 - y^2}\big) {\rm d}y \bigg) \\ \label{eq:pmf_den}
& \qquad \times \bigg[ \int_0^{y_n} \exp(-2 \lambda_v \sqrt{r^2 - y^2}) \frac{{\rm d}y }{y_n} \bigg]^{n-1},
\end{flalign}
where (a) follows from substituting \eqref{eq:pmf_num1} and \eqref{eq:pmf_num2} in \eqref{eq:pmf_den1}.

Substituting \eqref{eq:pmf_num1}, \eqref{eq:pmf_num2}, and \eqref{eq:pmf_den} in \eqref{eq:pnl_ren}, we obtain the final expression.

\subsection{Proof of Lemma \ref{lem:lap_Iin}}\label{app:lap_Iin}
We know that there are $n-1$ lines (excluding the typical line) closer than $Y_n$ whose distances from the origin are uniformly distributed in the range $(0, y_n)$. The Laplace transform of distribution of interference from an arbitrarily chosen line $L$, conditioned on its distance from the origin $Y$ is given by
\begin{align}\label{eq:lapIj_onelineY}
\calL_{I_{L}}(s | r, y_n, \calE_n, y ) = \exp\Bigg[-2 \lambda_v \bigintssss_{\sqrt{r^2 - y^2}}^{\infty} \bigg(1 -   \Big(1+ \frac{s (x^2 + y^2)^{-\alpha/2}}{m} \Big)^{-m} \bigg){\rm d}x \Bigg]. 
\end{align}
This expression is similar to the result obtained in Lemma \ref{lem:lapIn_En}. The lower limit of the integral follows from the condition that there must be no nodes closer than $r$. Now, taking the expectation over $Y$ which is uniformly distributed in the range $(0, y_n)$, we obtain the conditional Laplace transform of distribution of interference from a single line as follows:
\begin{align}\notag 
\calL_{I_{L}}(s | r, y_n, \calE_n ) &= \int_{0}^{y_n} \calL_{I_{L}}(s | r, y_n, \calE_n, y ) \frac{{\rm d}y}{y_n}\\
&= \bigintsss_0^{y_n} \exp\Bigg[-2 \lambda_v \bigintssss_{\sqrt{r^2 - y^2}}^{\infty} \bigg(1 -   \Big(1+ \frac{s (x^2 + y^2)^{-\alpha/2}}{m} \Big)^{-m} \bigg){\rm d}x \Bigg] \frac{{\rm d}y}{y_n}. 
\end{align}
Owing to the i.i.d. locations of nodes on the lines, the conditional Laplace transform of distribution of interference from the nodes on all the $n-1$ lines is given by
\begin{align}\notag
\calL_{I_{in} }&( s|r, y_n, \calE_n) = \big( \calL_{I_{L}}(s | r, y_n, \calE_n )\big) ^{n-1} \\
& = \Bigg( \bigintsss_0^{y_n} \exp\Bigg[-2 \lambda_v \bigintssss_{\sqrt{r^2 - y^2}}^{\infty} \bigg(1 -   \Big(1+ \frac{s (x^2 + y^2)^{-\alpha/2}}{m} \Big)^{-m} \bigg){\rm d}x \Bigg] \frac{{\rm d}y}{y_n}\Bigg)^{n-1}.
\end{align}

\subsection{Proof of Lemma \ref{lem:lapIin2_En}} \label{app:lap_Iann}
We know that the distances of the random number of lines that are farther than $y_n$ and closer than $r$ are uniformly distributed in the range $(y_n, r)$. As given in \eqref{eq:lapIj_onelineY}, the Laplace transform of interference from an arbitrarily chosen line $L$ conditioned on the distance of the line from the origin $Y$ is
\begin{align*}
\calL_{I_{L}}(s | r, y_n, \calE_n, y ) = \exp\Bigg[-2 \lambda_v \bigintssss_{\sqrt{r^2 - y^2}}^{\infty} \bigg(1 -   \Big(1+ \frac{s (x^2 + y^2)^{-\alpha/2}}{m} \Big)^{-m} \bigg){\rm d}x \Bigg]. 
\end{align*}
Thus, the conditional Laplace transform of distribution of interference from a single line is
\begin{align}\label{eq:lapIin2_step1}
\calL_{I_{L}}(s | r, y_n, \calE_n )  = \bigintsss_{y_n}^{r} \exp\Bigg[-2 \lambda_v \bigintssss_{\sqrt{r^2 - y^2}}^{\infty} \bigg(1 -   \Big(1+ \frac{s (x^2 + y^2)^{-\alpha/2}}{m} \Big)^{-m} \bigg){\rm d}x \Bigg] \frac{{\rm d}y}{(r-y_n)}. 
\end{align}
Owing to the i.i.d. locations of nodes on the lines, the conditional Laplace transform of interference distribution from all the lines that are farther than $Y_n$ and closer than $R$ can be computed as
\begin{align}\label{eq:lapIin2_step2}
\calL_{I_{ann} }(s|r, y_n, \calE_n) &= \sum_{i=0} ^{\infty} \P \Big(N_l\big(b(o, r)\setminus b(o, y_n)\big) = i\Big) \Big(\calL_{I_{L}}(s | r, y_n, \calE_n )\Big)^i\\ \label{eq:lapIin2_step3}
& \stackrel{(a)}{=} \exp\Bigg[\bigg(- 2 \pi \lambda_l \int_{y_n}^r\exp(-2 \lambda_v \sqrt{r^2-z^2}) {\rm d} z  \bigg) \bigg(1- \calL_{I_{L}}(s | r, y_n, \calE_n )\bigg)\Bigg],
\end{align}
where (a) follows from the substitution of the PMF given by Lemma \ref{lem:pmfNl_given_En} in \eqref{eq:lapIin2_step2}. Substituting \eqref{eq:lapIin2_step1} in \eqref{eq:lapIin2_step3}, we obtain the final expression.

\subsection{Proof of Lemma \ref{lem:lapIout_En}} \label{app:lap_Iout}
	Following the same approach as in Lemma \ref{lem:lapIin2_En}, we first determine the conditional Laplace transform of distribution of interference from an arbitrary line $L$ at a distance $Y$ from the origin as follows: 
\begin{align}\label{eq:lapIout_step1}
\calL_{I_{L}}(s | r, y_n, \calE_n, y ) = \exp\Bigg[-2 \lambda_v \bigintssss_{0}^{\infty} \bigg(1 -   \Big(1+ \frac{s (x^2 + y^2)^{-\alpha/2}}{m} \Big)^{-m} \bigg){\rm d}x \Bigg]. 
\end{align}
Owing to the independent distribution of nodes on the lines, for a given realization of the line process, the conditional Laplace transform of interference distribution is simply the product of the Laplace transform of distribution of interference from each of these lines. Therefore, we can write the Laplace transform of distribution of interference conditioned on the line process $\Phi_{l_0} \equiv \Phi_l \cup {L_0}$ as
\begin{align}
\calL_{I_{out}}(s | r, y_n, \calE_n, \Phi_{l_0} ) &= \prod_{y \in \Psi_{l_0}} \calL_{I_{L}}(s | r, y_n, \calE_n, y ),
\end{align}
where $\Psi_{l_0}$ represents the set of distances of the lines from the origin, which is a 1D PPP with density $2 \pi \lambda_l$ as shown in the proof of Lemma \ref{lem:cdf_Yn}.
By taking the expectation over $\Psi_{l_0}$, we obtain the desired result as follows:
\begin{align}\notag 
\calL_{I_{out}} (s | r, y_n, \calE_n) &= \E_{\Psi_{l_0}} \bigg[\prod_{y \in \Psi_{l_0}} \calL_{I_{L}}(s | r, y_n, \calE_n, y ) \bigg] \\ \label{eq:lapIout_step2}
&\stackrel{(a)}{=} \exp\bigg[ -2 \pi \lambda_l \int_r^{\infty} \big(1 - \calL_{I_{L}}(s| r, y_n, \calE_n, y) \big){\rm d}y \bigg],
\end{align}
where (a) follows from the PGFL of PPP $\Psi_{l_0}$. We obtain the final expression upon substituting \eqref{eq:lapIout_step1} in \eqref{eq:lapIout_step2}.

\bibliographystyle{IEEEtran}

\bibliography{j2_v0.16.bbl}

\end{document}